\documentclass[%
 reprint,
superscriptaddress,https://www.overleaf.com/project/64994f20bd1198e1d7691150
 amsmath,amssymb,
 aps,pre,hyphens,floatfix
]{revtex4-1}

\usepackage{lineno}
\usepackage{subfigure}
\usepackage{parskip}

\usepackage{hyperref}
\hypersetup{hypertex=true,
colorlinks=true,
linkcolor=blue,
anchorcolor=blue,
citecolor=blue}
\usepackage{enumerate}
\usepackage[usenames,dvipsnames,svgnames,table]{xcolor}
\usepackage{IEEEtrantools}
\usepackage{graphicx}
\usepackage{array}
\usepackage{multirow}
\usepackage{verbatim}
\usepackage{float}
\usepackage{amsmath}

\begin{document}
\preprint{APS/123-QED}

\title{Autoencoder-assisted study of directed percolation with spatial long-range interactions}

\author{Yanyang Wang}
\affiliation{Key Laboratory of Quark and Lepton Physics (MOE) and Institute of Particle Physics, Central China Normal University, Wuhan 430079, China}
\author{Yuxiang Yang}
\affiliation{Key Laboratory of Quark and Lepton Physics (MOE) and Institute of Particle Physics, Central China Normal University, Wuhan 430079, China}
\author{Wei Li}
\affiliation{Key Laboratory of Quark and Lepton Physics (MOE) and Institute of Particle Physics, Central China Normal University, Wuhan 430079, China}



\begin{abstract}

In the field of non-equilibrium phase transitions, determining the universality class of reaction-diffusion processes with long-range interactions is both challenging and intriguing. Identifying critical points serves as the foundation for studying the phase transition characteristics of these universality classes. In contrast to Monte Carlo simulations of statistical system observables, machine learning methods can extract evolutionary information from clusters of such systems, thereby rapidly approaching phase transition regions. We have developed a method to determine critical points in phase transition systems by utilizing the one-dimensional encoded outputs from stacked autoencoders. We validate this method by examining the power-law behavior of particle density at the critical point, which lends strong credibility to our approach. Given that the system adheres to the scaling relation $t_f{\sim}L^{z}$ at the critical point, we performed extensive simulations at this critical probability to ascertain the dynamic exponent $z$. In addition, the stacked autoencoder is also capable of identifying the characteristic time of critical states. Finally, we tested an alternative approach adopting L{\'{e}}vy distribution to generate random walk steps, inserting another global expansion mechanism. This method remains effective for determining critical points. Our findings highlight the promising applications of stacked autoencoder techniques in processes involving long-range interactions.



\end{abstract}
\maketitle



\section{Introduction}
\label{intro}


In recent years, machine learning has been widely applied in various fields of physics\cite{goodfellow2016deep}. With the powerful mapping and generalization capabilities of neural networks, both supervised and unsupervised learning have found numerous applications. Examples include astronomy\cite{huerta2019enabling,0Probabilistic,smith2023astronomia}, quantum information\cite{ma2018transforming,kookani2023xpookynet,Zhang_2021}, high-energy physics\cite{erbin2022characterizing,Ma2022AJT,steinheimer2019machine}, biophysics\cite{jian2017direct,yan2020revealing,tareen2019biophysical,giulini2019deep}, and complexity science\cite{maseer2023meta,xie2023simple}.The research methods related to complexity and critical phenomena include but are not limited to, mean-field theory, renormalization, exact diagonalization \cite{christensen2005complexity,2005Field}, and numerical simulation methods such as Monte Carlo (MC) simulations\cite{1986Monte,lubeck2006crossover,hinrichsen2000non}. Machine learning can be involved in both the theoretical and numerical solution processes of phase transition models, greatly enriching the solution approaches and the scope of their applications. For equilibrium systems, statistical methods based on general ensemble theory are well established\cite{henkel2008non}. However, the prevalence of open systems in nature forces us to consider the dynamical behavior of non-equilibrium systems. The dialectical relationship between equilibrium and non-equilibrium systems informs us about the applicability of theoretical methods such as mean-field theory and field theory renormalization in non-equilibrium phase transition models\cite{jensen2004low,munoz1996critical,cardy1998field,cardy1996theory}. It also inspires us to explore the possibility of combining MC simulations and machine learning for numerical solutions\cite{carrasquilla2017machine}.


Machine learning has made significant progress in some equilibrium phase transition models\cite{Chen2022StudyOP,wang2016discovering,hu2017discovering}. For non-equilibrium phase transitions, the absence of detailed balance allows for richer critical behavior in systems that are far from equilibrium. Absorbing phase transitions are a class of continuous phase transitions in non-equilibrium systems, where the transition occurs between an absorbing state with no surviving particles and an active state with active particles, controlled by a series of reaction-diffusion processes in the particle dynamics evolution. An important universality class of absorbing phase transitions is the directed percolation(DP) university class, characterized by consistent critical exponents and exemplified by the DP model. The measurement of a series of critical exponents is a vital reference for determining the universality class to which a model belongs. Within the theoretical approaches, the DP universality is identified by several fundamental conditions\cite{1981On,1982On}. In a broader context of particle reaction-diffusion, the DP universality class is observed in branching-annihilating random walks(BAW) processes with an odd number of branches\cite{1992Extinction,1994Critical}. Conversely, the BAW models with an even number of branches belong to the parity-conserving(PC) universality class\cite{Menyhard1994Domain,zhong1995universality,canet2005nonperturbative}. The use of supervised and unsupervised machine learning methods to study the critical properties of non-equilibrium phase transition models appears to offer promising applications and research potential\cite{shen2021machine,shen2022transfer,Shen2021SupervisedAU}. Among these approaches, unsupervised learning methods provide a way to extract features near the critical point of a system\cite{stoudenmire2018learning,bro2014principal,wattenberg2016use,wang2016auto}.


One of the conditions for ensuring the robustness of the directed percolation (DP) universality class is to guarantee that the system only exhibits local interactions in both time and space, which aligns with the consideration of reactions and diffusion involving only nearest-neighbor particles in processes such as DP and BAW model. However, when considering the coupling of interactions and potential distributions in a lattice model, bringing in long-range interactions into reaction-diffusion systems can better reflect real physical systems. Taking into account L\'evy-like flights in DP may alter its universality class, and it may find broader applications under conditions such as long-range infection, latent periods, and memory effects in realistic scenarios\cite{2000Stochastic}. In the context of epidemic spread, Mollison proposed an extension of directed percolation with non-local spreading mechanisms\cite{1977Spatial}, where diseases spread over a distance of $r$  in a $d$-dimensional space, with $r$ following a typical power-law distribution
\begin{equation}
 P(r){\sim}{\frac{1}{r^{d+{\sigma}}}}{\sim}{\frac{1}{r^{\beta}}}.
\label{1}
\end{equation} 
The random walk displacement that satisfies this algebraic distribution is known as L\'evy-like flights \cite{1990Anomalous}. Levy-like flights have a shorter time scale compared to the nearest-neighbor propagation models, resulting in non-local effects and longer distance extensions. In the analysis of probability density evolution for particle random walk models, L\'evy flights can be generated by introducing nonlinear operators, also known as fractional order derivatives\cite{Hinrichsen_2007}. In our numerical simulation approach, we incorporate long-range interactions in the reaction-diffusion process through the settings of random numbers and step sizes. We discuss the relevant simulation details of introducing L\'evy-like flights into the DP model at the spatial scale, predict critical points, and measure several critical exponents, based on the the DP model with spatial long-range interactions.


Utilizing unsupervised learning to identify and predict the structural characteristics of the evolution of phase transition models is one of the fundamental methods in applying machine learning to study phase transition\cite{shen2021machine}. Stacked autoencoders (SAE), which combine fully connected neural networks and autoencoders, are one type of unsupervised learning algorithm. The primary objective of an autoencoder, involving an encoder, decoder, and loss function, is data dimensionality reduction and reconstruction. The learning process of an autoencoder can be regarded as the minimization of a loss function. Fully connected neural networks, on the other hand, provide a data compression method when dealing with grid-like structured data. When employed as a supervised learning algorithm, fully connected neural networks can effectively identify critical states of some phase transition models\cite{carrasquilla2017machine}. 
The basic structure of SAE involves gradually stacking fully connected layers in the encoding and decoding processes. In practice, the structural details of SAE often need adjusting according to the system size.
We consider the potential of utilizing SAE in the encoding process to identify critical states of spatial L\'evy-like flights in the DP process, aiming to explore the feasibility of applying unsupervised learning methods based on autoencoders to study the critical properties of long-range interaction non-equilibrium phase transition models in the context of 
(1+1)-dimensional spatial L\'evy-like flights DP (LDP) models.


The structure of this paper is as follows: In Sec.\uppercase\expandafter{\romannumeral2}, We briefly introduced the manifestation of long-range interactions in the theoretical LDP model.
In Sec.\uppercase\expandafter{\romannumeral3}.A, we discuss the simulation details of introducing spatial long-range interactions into the DP model and present some numerical simulation results of the evolution. 
In Sec.\uppercase\expandafter{\romannumeral3}.B, we outline the general process of SAE methods and discuss how certain settings affect the training process. 
Sec.\uppercase\expandafter{\romannumeral4}.A  provides a series of predicted critical points based on the one-dimensional encoding output using SAE. 
In Sec.\uppercase\expandafter{\romannumeral4}.B, we observe the decay behavior of the system's particle density at these critical points to determine the critical exponent ${\delta}$. 
Furthermore, in Sec.\uppercase\expandafter{\romannumeral4}.C, we investigate the growth of active particles at critical points to determine the characteristic time $t_f$ of finite-scale systems, thereby obtaining the measured value of the dynamic exponent $z$. SAE can effectively identify these characteristic times. 
Finally, in Sec.\uppercase\expandafter{\romannumeral4}.D, We studied a new method for generating random walk step lengths to test the universality of our approach.
In Sec.\uppercase\expandafter{\romannumeral5}, we summarize this work and provide an outlook on future research directions.

\section{The Model of DP with spatial levy-like flights}


In the framework of particle reaction-diffusion, the continuous phase transition from an active state to an inactive state in a system demonstrates that the dynamic evolution of such absorbing phase transitions is truly a non-equilibrium process influenced by fluctuations. In order to investigate the probability distribution analysis of non-equilibrium system structures, it is necessary to abandon the detailed balance condition and the Einstein relation that controls the long-time evolution direction of the system in setting up the dynamics equations for such systems \cite{tauber2014critical,1989Quantum,2015Field}. 

\begin{figure*}[t]
    \centering
        \includegraphics[width=0.8\textwidth]{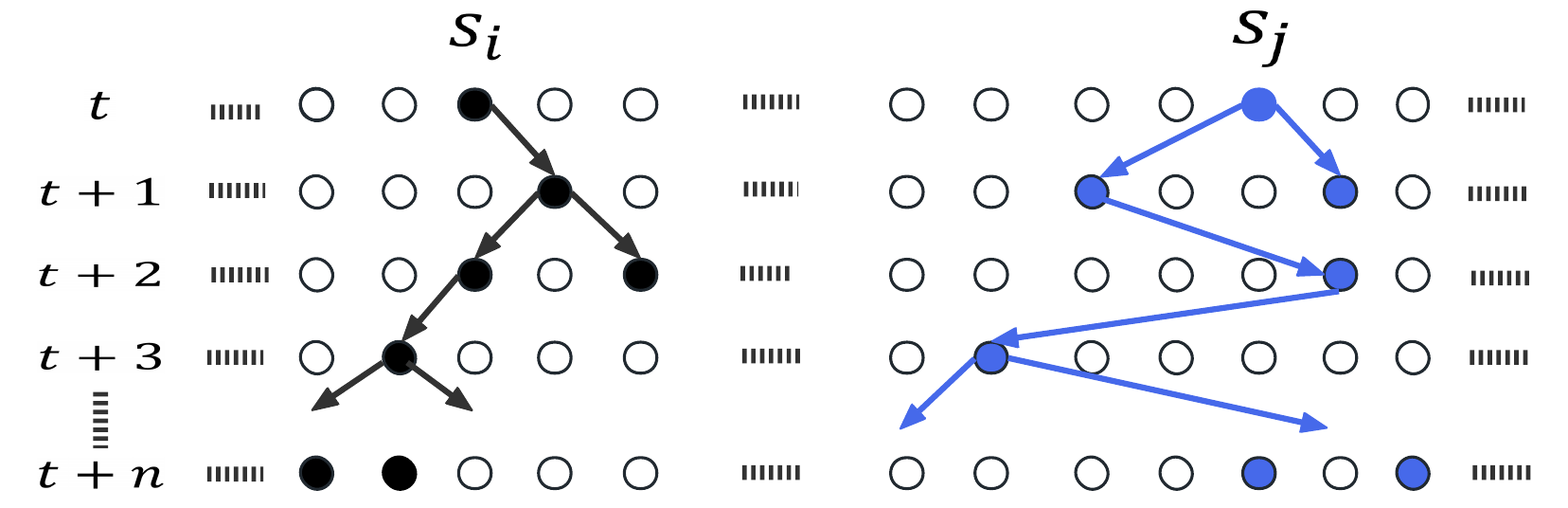} 
\caption
{The evolution of (1+1)-dimensional ordinary DP and LDP. In the following illustrations, black and blue dots represent occupied sites, while empty dots indicate unoccupied sites. The left panel depicts the evolution rules (\ref{8}) for ordinary DP, with black particles evolving over time steps starting from $S_i$ following the black arrows. The blue dots represent an example of the evolution of LDP at $S_j$ according to rules (\ref{9}). Unlike ordinary DP, the interaction range of LDP is not limited to nearest neighbors, allowing particles to appear further apart in a shorter period of time.}
\label{f1}
\end{figure*}

\begin{figure*}[t]
\begin{tabular}{cc}
    \includegraphics[width=0.46\textwidth]{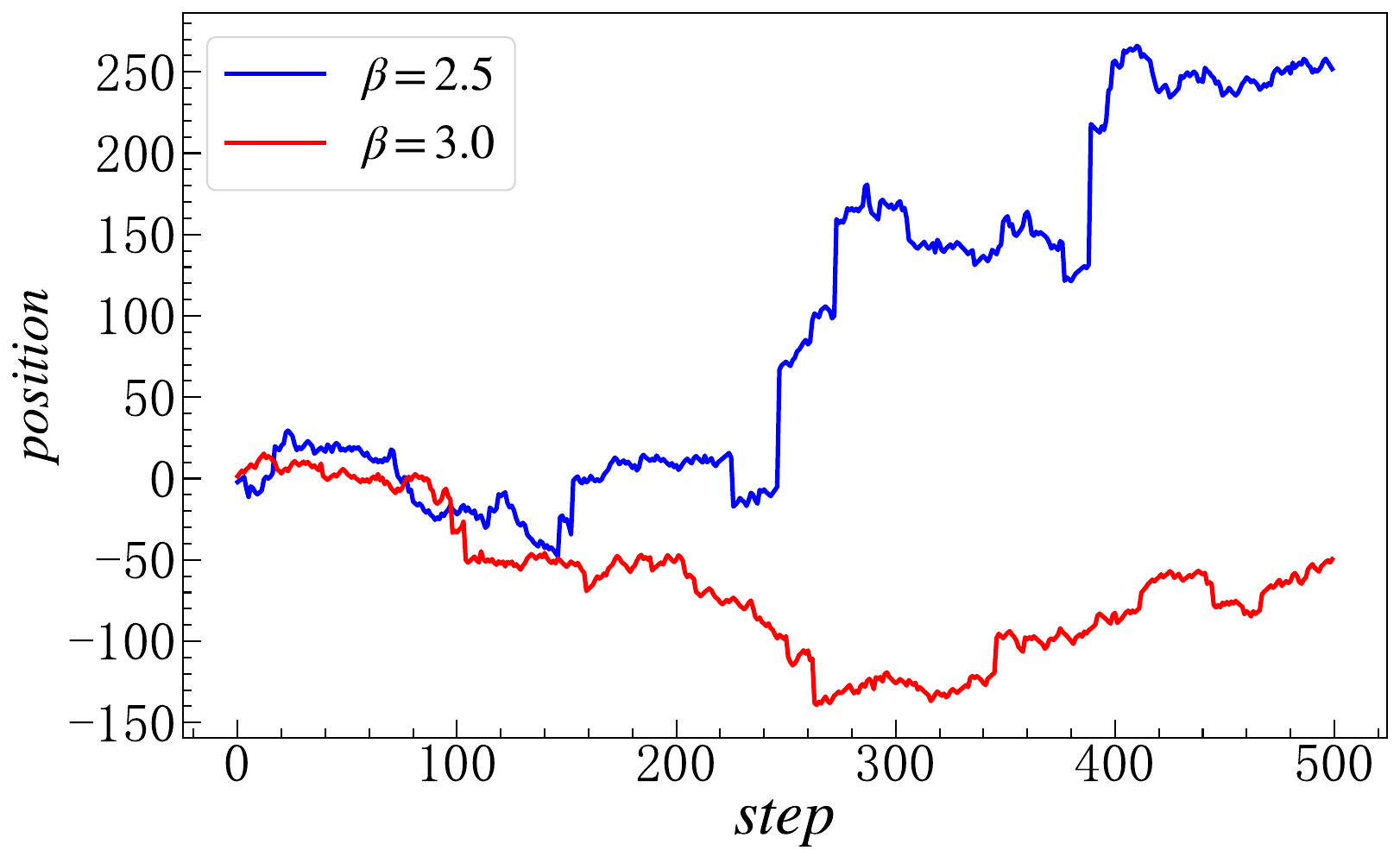} &
    $\qquad$\includegraphics[width=0.46\textwidth]{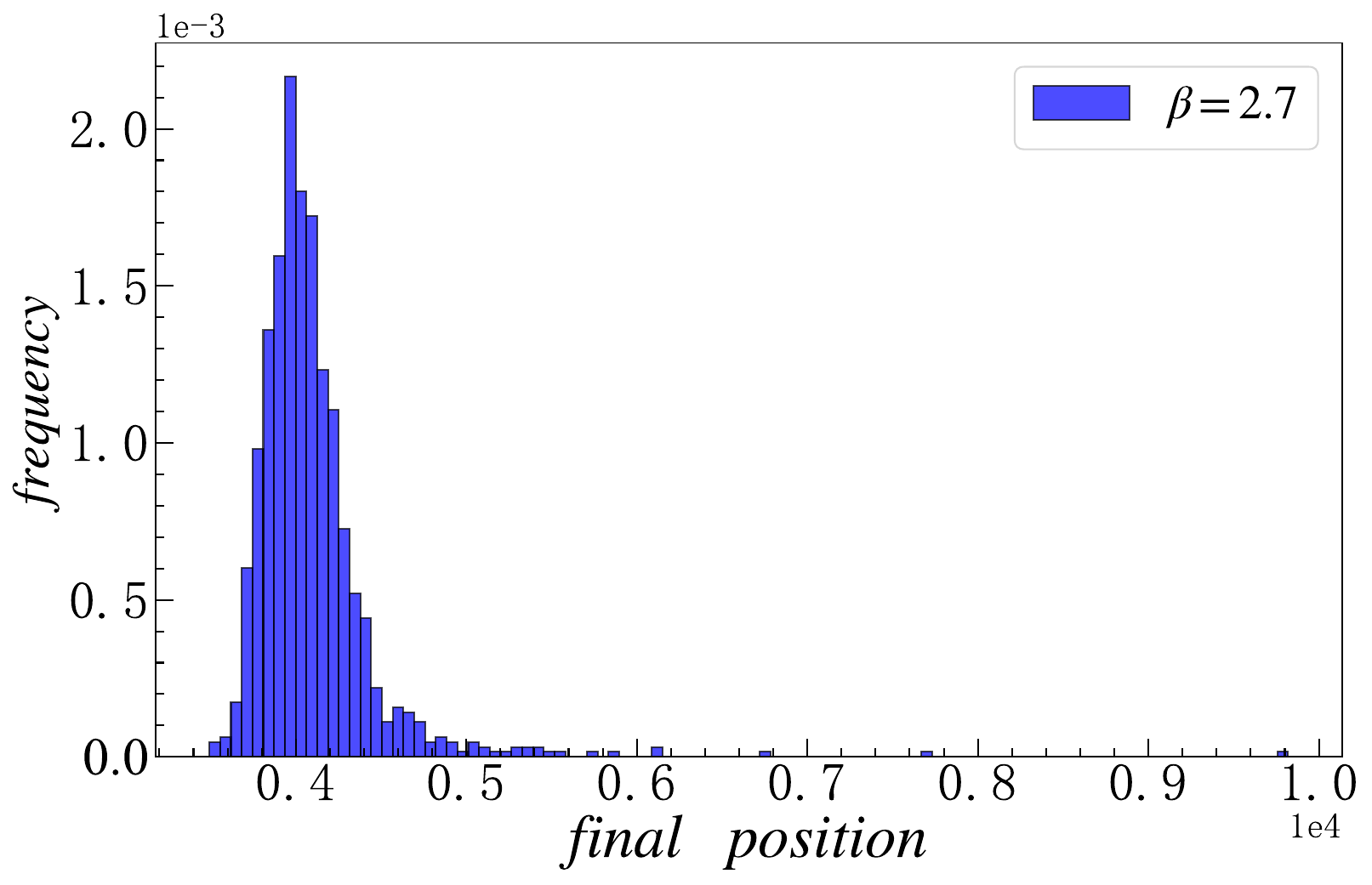} \\
    {\quad}{\quad}(a) & $\qquad$ {\quad}{\quad}(b)
\end{tabular}
\caption
{(a) A one-dimensional random walk conducted based on the step sizes generated according to formula (\ref{10}) for ${\beta}=2.5,=3.0$. The horizontal axis represents the number of steps in the random walk, while the vertical axis indicates the position of the particle at the current step. It can be observed from the graph that larger step sizes can be generated for smaller ${\beta}$ values. In (b), with the number of steps set to 2000, a distribution plot of the final positions of particles after 1000 independent random walks is shown. The horizontal axis represents the final positions where particles appear, and the vertical axis represents the corresponding frequency of those positions. It is evident from the graph that there are pronounced characteristics of a long-tailed distribution.}
\label{f2}
\end{figure*}




\begin{figure*}[t]
\begin{tabular}{cc}
    \includegraphics[width=0.30\textwidth]{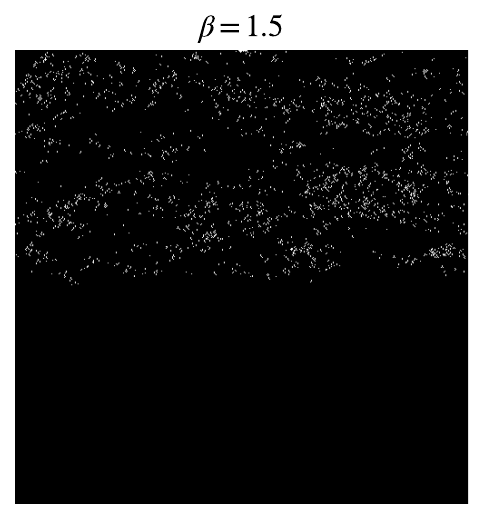} &
    $\qquad$\includegraphics[width=0.30\textwidth]{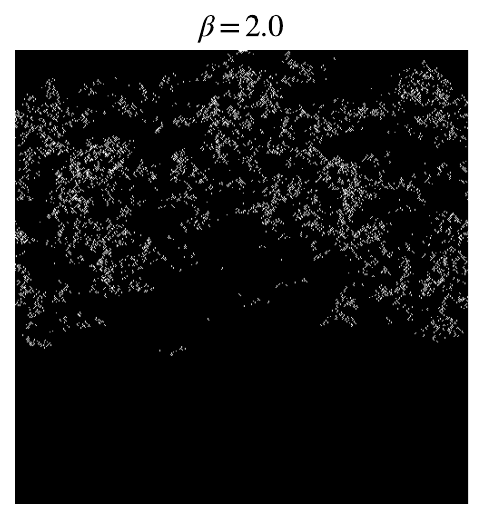} \\
    {\quad}{\quad}(a) & $\qquad$ {\quad}{\quad}(b)
\end{tabular}
\begin{tabular}{cc}
    \includegraphics[width=0.30\textwidth]{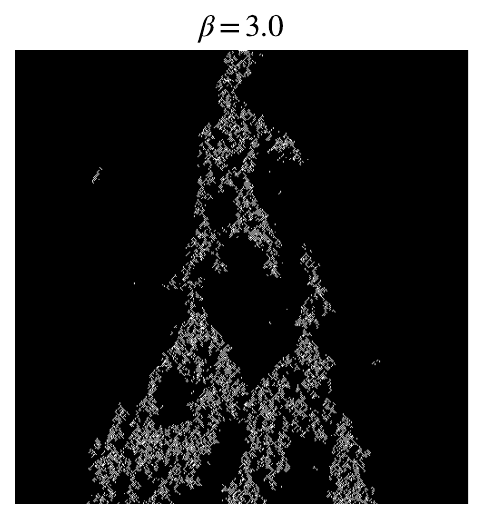} &
    $\qquad$\includegraphics[width=0.30\textwidth]{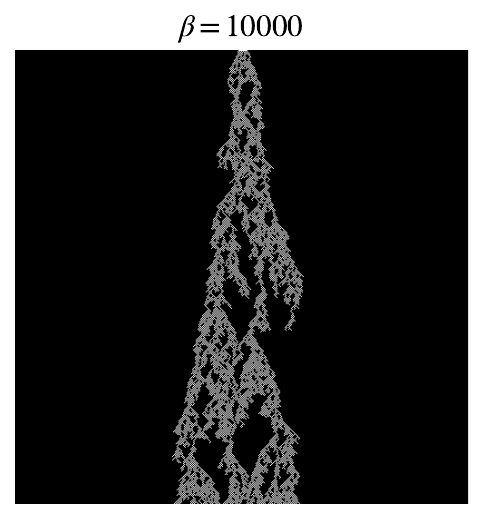} \\
    {\quad}{\quad}(c) & $\qquad$ {\quad}{\quad}(d)
\end{tabular}
\caption
{The clusters growth structure starting from an initial set of 10 active seeds for ${\beta}=1.5,2,3,10000$, with a system size of $L=500$ and a time step setting of $t=500$. When ${\beta}$ is small, the system may transition to the absorbing state more rapidly, indicating a decrease in the characteristic time of system evolution, and clusters tend to become more dispersed. As ${\beta}$ increases, the system evolution leads to the formation of larger clusters and a more ordered structure. When ${\beta}=10000$, the cluster growth structure closely resembles the evolution of ordinary DP.}
\label{f3}
\end{figure*}

Below the upper critical dimension, it is necessary to fully consider the system's fluctuation effects. Field-theoretic renormalization group (RG) methods can provide explanations of symmetry and universality for such reaction-diffusion systems, but it is not particularly good at accurately predicting the critical points and the critical exponents.The renormalization analysis of the DP model with spatial L\'evy-like flights is based on the construction of the action\cite{1981On}\cite{Janssen_1999}:
\begin{equation}
\begin{gathered}
S[\bar{\psi}, \psi]=\int d^d x d t\left[\bar{\psi}\left(\partial_t-\tau-D_N \nabla^2-D_A \nabla^\sigma\right) \psi +\right. \\
\left.+\frac{g}{2}\left(\bar{\psi} \psi^2-\bar{\psi}^2 \psi\right)\right] .
\end{gathered}
\label{5}
\end{equation}
where $\nabla^{\sigma}$ describes the non-local reaction-diffusion behaviour, and the operator $\nabla^{\sigma}$ is known as a fractional-order derivative, characterized by its properties as follows\cite{Janssen_1999}:
\begin{equation}
\nabla^{\sigma} \mathrm{e}^{\mathrm{i} \vec{k} \cdot \vec{r}}=-|\vec{k}|^\sigma \mathrm{e}^{\mathrm{i} \vec{k} \cdot \vec{r}}.
\label{4}
\end{equation}


In numerical simulations, long-range interactions can be modeled by modifying the reaction-diffusion distance between lattice points. Unlike the continuous operator $\nabla^{\sigma}$ used in theoretical studies, the interaction distance between particles on a lattice is typically discrete. Varying the rules for random step lengths can result in different outcomes regarding long-range interactions. In the following section, we will present a method for altering the characteristics of ordinary directed percolation by incorporating global interactions in space.

\section{Model and Autoencoder method}
\subsection{Simulation of the DP with spatial L{\'{e}}vy-like flights}




By setting the transition probabilities within the Domany-Kinzel automaton (DK)\cite{henkel2008non,1984Equivalence,1985Phase}, the update rules for the ordinary (1+1)-dimensional DP process can be determined. The basic setup of the DK cellular automaton model involves using the occupation status of surrounding lattice points to determine the occupation state of a lattice point at the next time step. For the ordinary DP, the occupation state of the point $s_{i,t}$ depends only on its nearest neighbors $s_{i-1,t}$ and $s_{i+1,t}$. The update rules for bond DP can be expressed as: 

\begin{equation}
s_{i, t+1}= 
\begin{cases}
1 & \text { if } s_{i-1, t} \neq s_{i+1, t} \quad \text { and } z_i(t)<p ,\\ 
1 & \text { if } s_{i-1, t}=s_{i+1, t}=1 \text { and } z_i(t)<p(2-p), \\ 
0 & \text { otherwise },
\end{cases} \label{8}
\end{equation}

where $s_{i,t+1}=1$ represents a site being occupied, and $s_{i,t+1}=0$ represents a site not being occupied. $z_i(t)$ is a uniformly distributed random number in the interval $[0,1]$, and $p$ is a hyperparameter representing the transition probability, often controlling the transition between the absorbing and active states.




A general way to introduce the space L{\'{e}}vy-like flights into the ordinary DP process described above is to change the influences on the state occupied by the locus at the next moment. This entails replacing the nearest neighbors $s_{i-1,t}$ and $s_{i+1,t}$ with $s_{i-[L],t}$ and $s_{i+[R],t}$, where $[L]$ and $[R]$ represent the largest positive integers not exceeding the distances $L$ and $R$, respectively. In this case, the update rules for LDP can be expressed as:
\begin{equation}
s_{i, t+1}= 
\begin{cases}
1 & \text { if } s_{i-[L], t} \neq s_{i+[R], t} \quad \text { and } z_i(t)<p ,\\ 1 & \text { if } s_{i-[L], t}=s_{i+[R], t}=1 \text { and } z_i(t)<p(2-p), \\ 0 & \text { otherwise }.
\end{cases} \label{9}
\end{equation}



When setting the generation rules for $L$ and $R$, different forms of spatial long-range interactions can be introduced. Figure \ref{f1} represents the evolution of ordinary DP and LDP, where the percolation action of the blue points is not limited to the nearest neighboring lattice points. We are considering the spatial long-range interactions that follow power-law distributions. It's worth noting that there are multiple methods for generating random walk step sizes that satisfy power-law distributions, and in this part, we have employed 
\begin{equation}
    \begin{aligned}
    &[L]=Max(L),L=Z_L^{-1/({\beta}-1)},
    \\
    &[R]=Max(R),R=Z_R^{-1/({\beta}-1)}.  
    \end{aligned} \label{10}
\end{equation} 
to define the generation rule of step size. Here, the function $Max(L)$ denotes the maximum integer not exceeding $L$, while $Z_L,Z_R\in(0,1)$ are random numbers following a uniform distribution. $\beta$ is a positive real number greater than $1$. 






\begin{figure*}[t]
    \centering
        \includegraphics[width=0.8\textwidth]{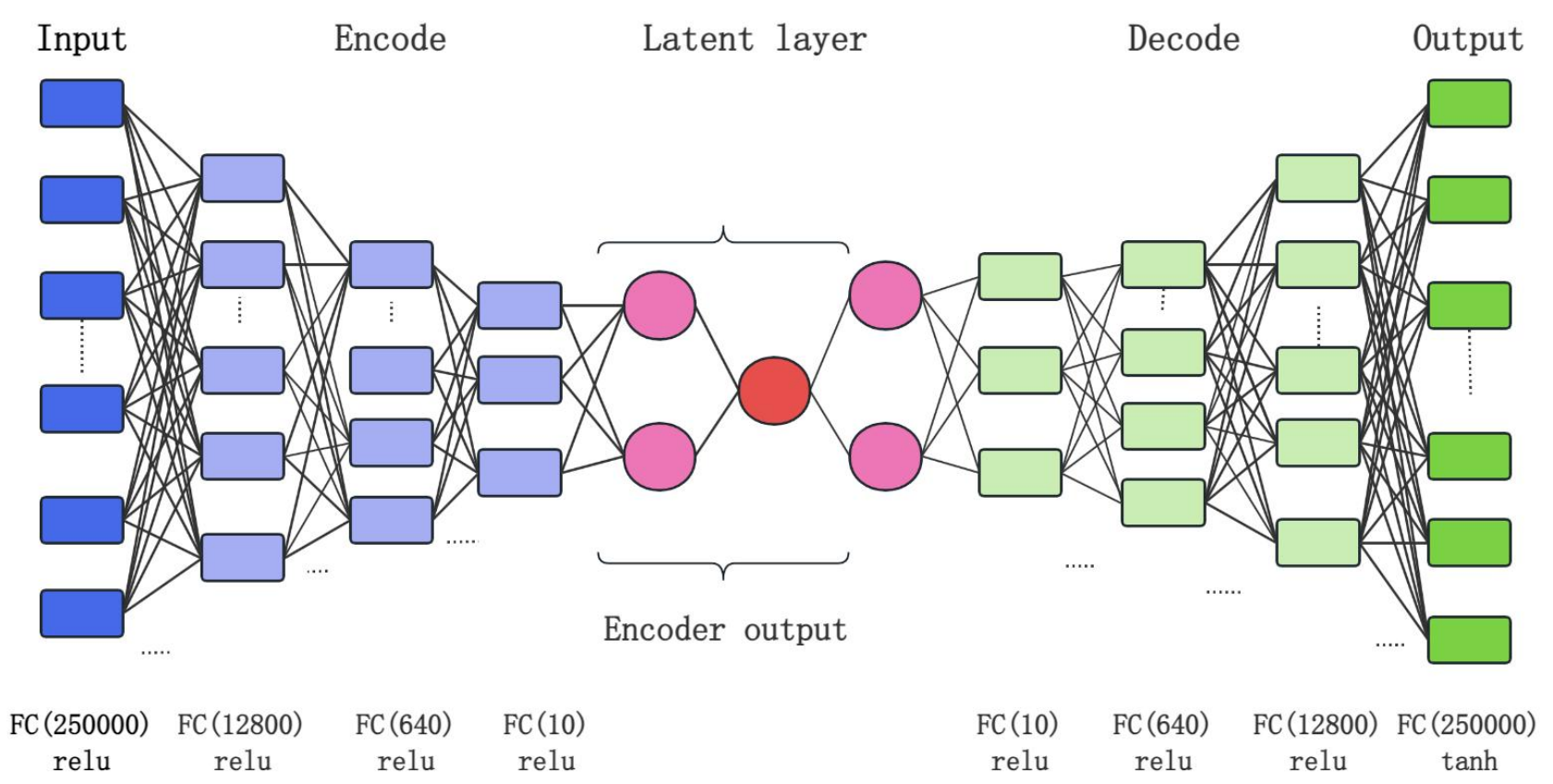} 
\caption
{The general structure of an SAE is designed with layered hidden units to preserve the original cluster graph information as much as possible. The Encoder output section in the figure is extracted after the training of the SAE. Two pink-colored neurons are used to analyze structural features, while one red-colored neuron is used to determine the critical point.}
\label{f_3_1}
\end{figure*} 

At ${\beta}=2.0,3.0$, we generate 500 steps for one-dimensional random wandering, as shown in Figure \ref{f2}(a). From the figure, it can be observed that as $\beta$ is smaller, it is possible to generate larger step sizes. In addition, Figure \ref{f2}(b) shows the distribution of step lengths for $2000$ random walks, from which an obvious "long-tailed distribution" can be seen. The diffusion of particles from the site $s_i$ at time $t$ to the site at time $t+1$ depends on the generation of step sizes $[L]$ and $[R]$. We employ the $Max$ function to ensure that interactions cover the entire lattice. For example, particles at position $s_i$ can diffuse to$s_{i-[L]},s_{i-[L-1]},...,s_{i},s_{i+1},s_{i+2},...,s_{i+[R-1]},s_{i+[R]}$.

We implemented a simulation program in Python3.7 to simulate the evolution rules of LDP, utilizing periodic boundary conditions to reduce finite-size effects. In Figure \ref{f3}, we present the growth results of clusters under different $\beta$ values. When $\beta$ is small, the system is more likely to quickly enter an absorbing state, indicating a smaller characteristic time and a tendency for clusters to disperse. As $\beta$ increases, the system forms larger clusters with a more ordered structure. 
From the perspective of system fluctuations, increasing $\beta$ generally amplifies the effects of fluctuations, resulting in a rise in the upper critical dimension. 

\subsection{Method of Stacked Autoencoder}

\begin{figure}
\centering
\includegraphics[width=0.45\textwidth]{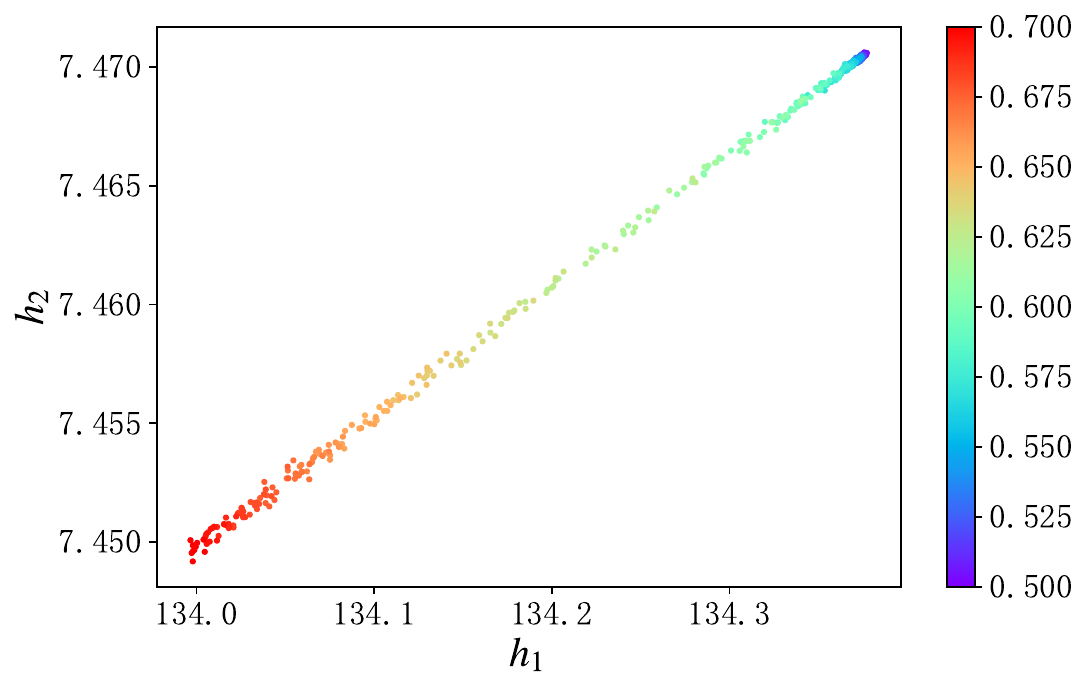}
\caption{Two-dimensional feature extraction of \;$(1+1)$\;-dimensional LDP training by SAE.The right color bars represent different values of directed transition probabilities. The scatter plot of the two-dimensional encoding of the stacked autoencoder shows a dense clustering of points far from a certain transition probability, while the points become sparser as they approach this probability.}
\label{f_4}
\end{figure}

Different from traditional supervised learning methods, unsupervised learning may be capable of predicting the critical properties of absorbing phase transitions. In the absence of any prior information about the system's dynamic evolution, the objective of unsupervised learning can be to provide the probability distribution of random vectors \cite{bro2014principal,wattenberg2016use}. Among these methods, autoencoders, as a mature unsupervised learning approach, have been applied in the study of phase transitions and critical phenomena. An autoencoder is a type of neural network with the fundamental training objective of attempting to replicate the input to the output or perform incomplete input replication. By setting different loss functions, the encoding and decoding effectiveness of the autoencoder can be evaluated. Considering the effectiveness of autoencoders in handling image data, we attempted to use an autoencoder with a fully connected neural network structure to process cluster configuration of absorbing phase transitions. Given the particular LDP model near the critical point, our basic approach is to utilize the dimensionality reduction function of the autoencoder to extract the spatial and temporal structural characteristics of the system's cluster configurations and compare the characteristic outputs under different transition probabilities to determine the position of the critical point.

\begin{figure*}[t]
\begin{tabular}{cc}
    \includegraphics[width=0.47\textwidth]{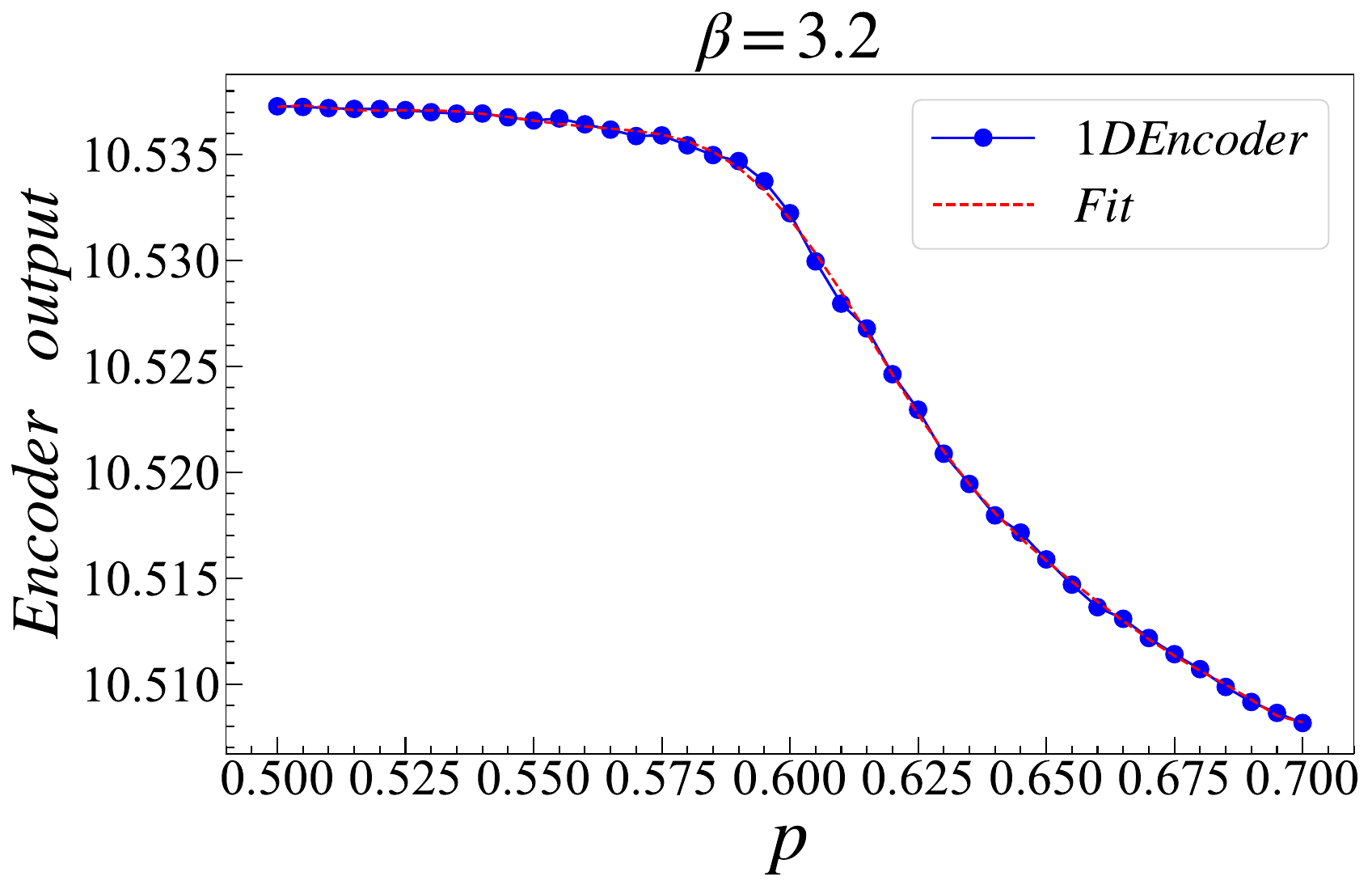} &
    $\qquad$\includegraphics[width=0.45\textwidth]{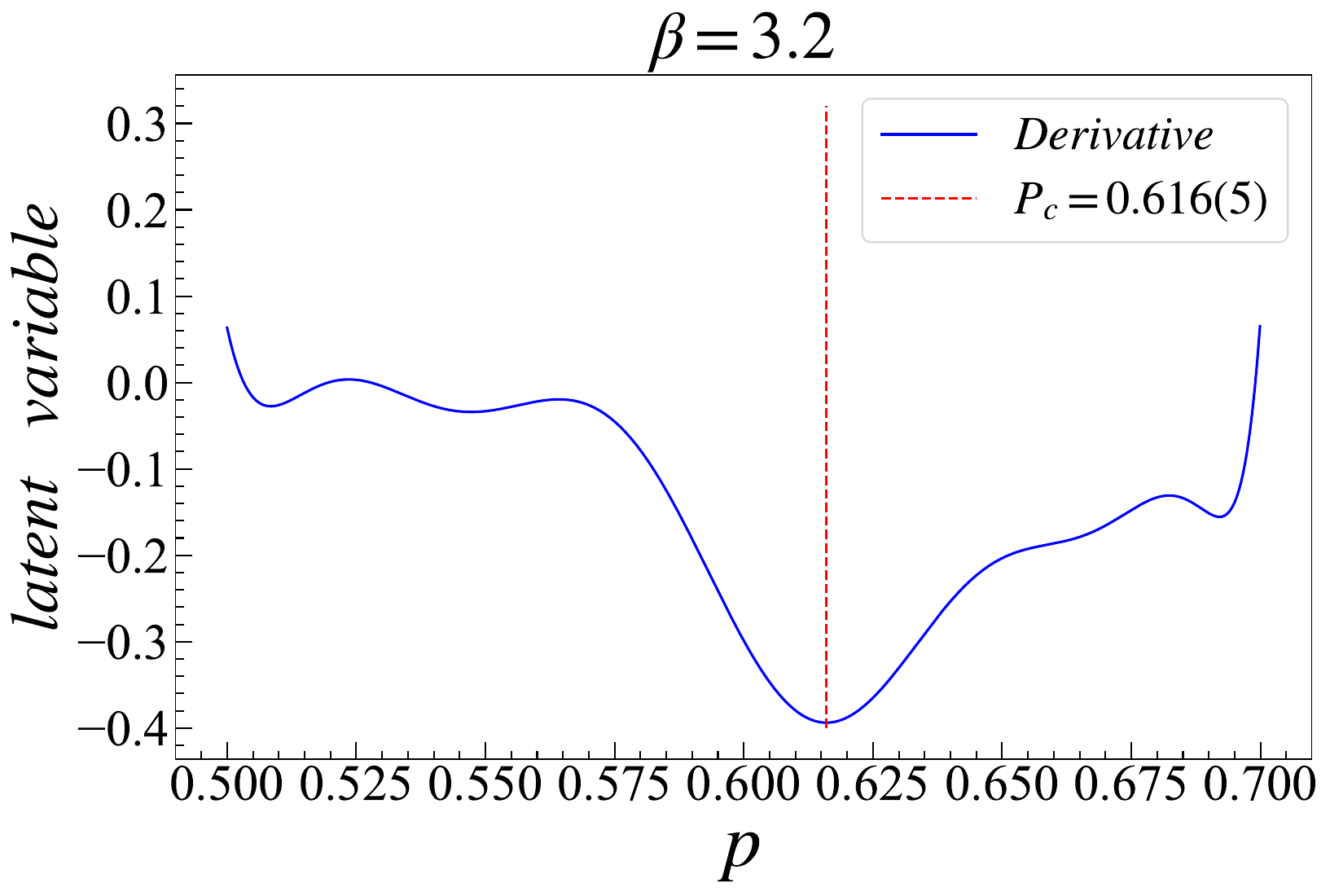} \\
    {\quad}{\quad}{\quad}(a) & $\qquad$ {\quad}{\quad}(b)
\end{tabular}
\caption
{At ${\beta}=3.2$, the one-dimensional encoding output of the hidden layer in SAEs, and the determination of the LDP critical point $P_c$. (a) For a system size of $500$ and a time step of $500$, we obtained the one-dimensional output of SAEs for $41$ transition probability $p$ in the range of $[0.50, 0.70]$. The blue dots represent the one-dimensional encoding output of SAEs, while the red dashed line represents the result of polynomial fitting. Under the settings of Figure (a), the derivative of the fitted result of the one-dimensional encoding output of SAEs is shown by the blue curve in (b). By identifying the global minimum in the extreme value of the derivative curve, we determined the position of the system's critical point characterized by the transition probability $P_c=0.616(5)$.}
\label{f_5}
\end{figure*}

\begin{figure*}[t]
\begin{tabular}{cc}
    \includegraphics[width=0.47\textwidth]{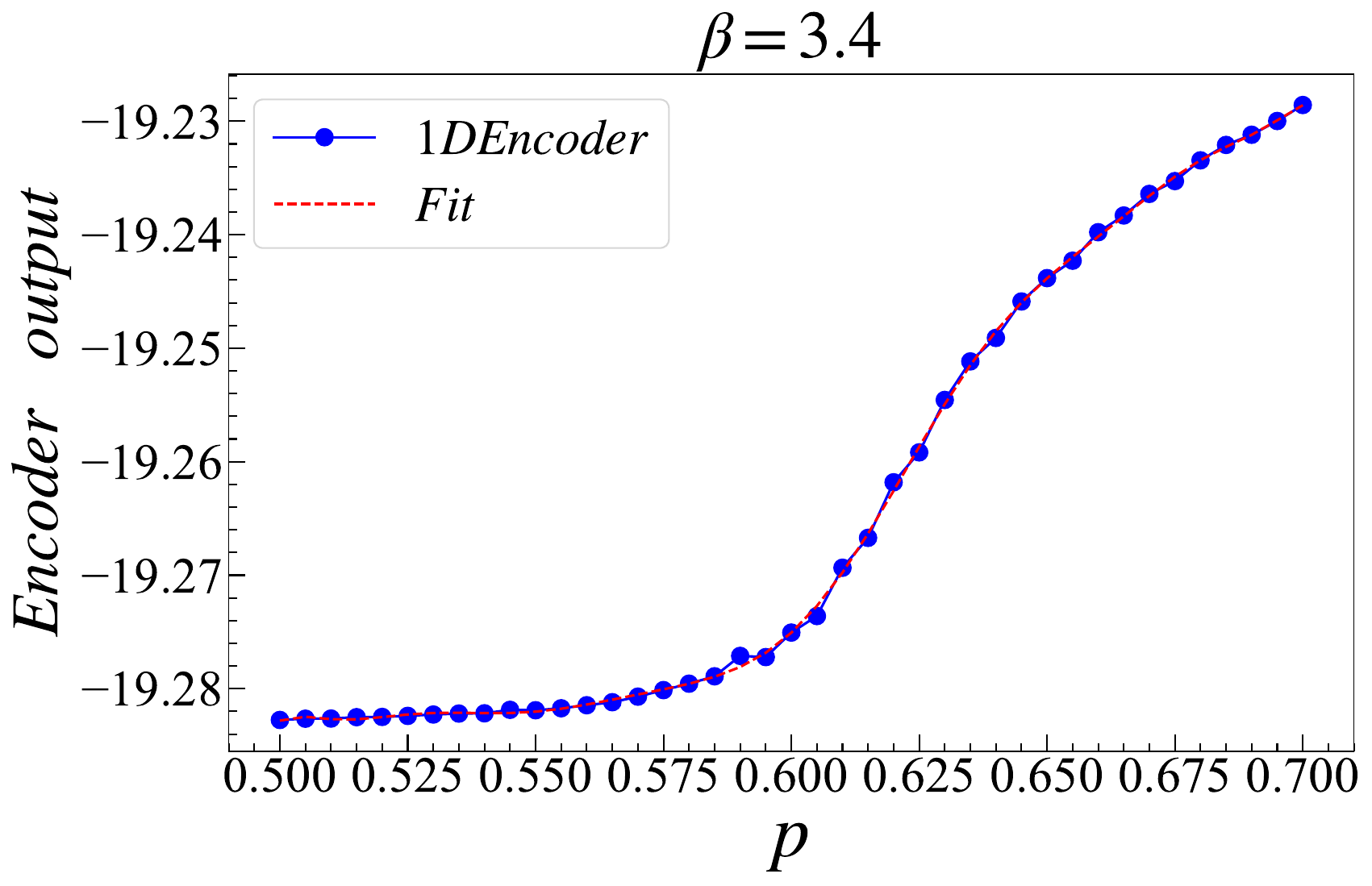} &
    $\qquad$\includegraphics[width=0.45\textwidth]{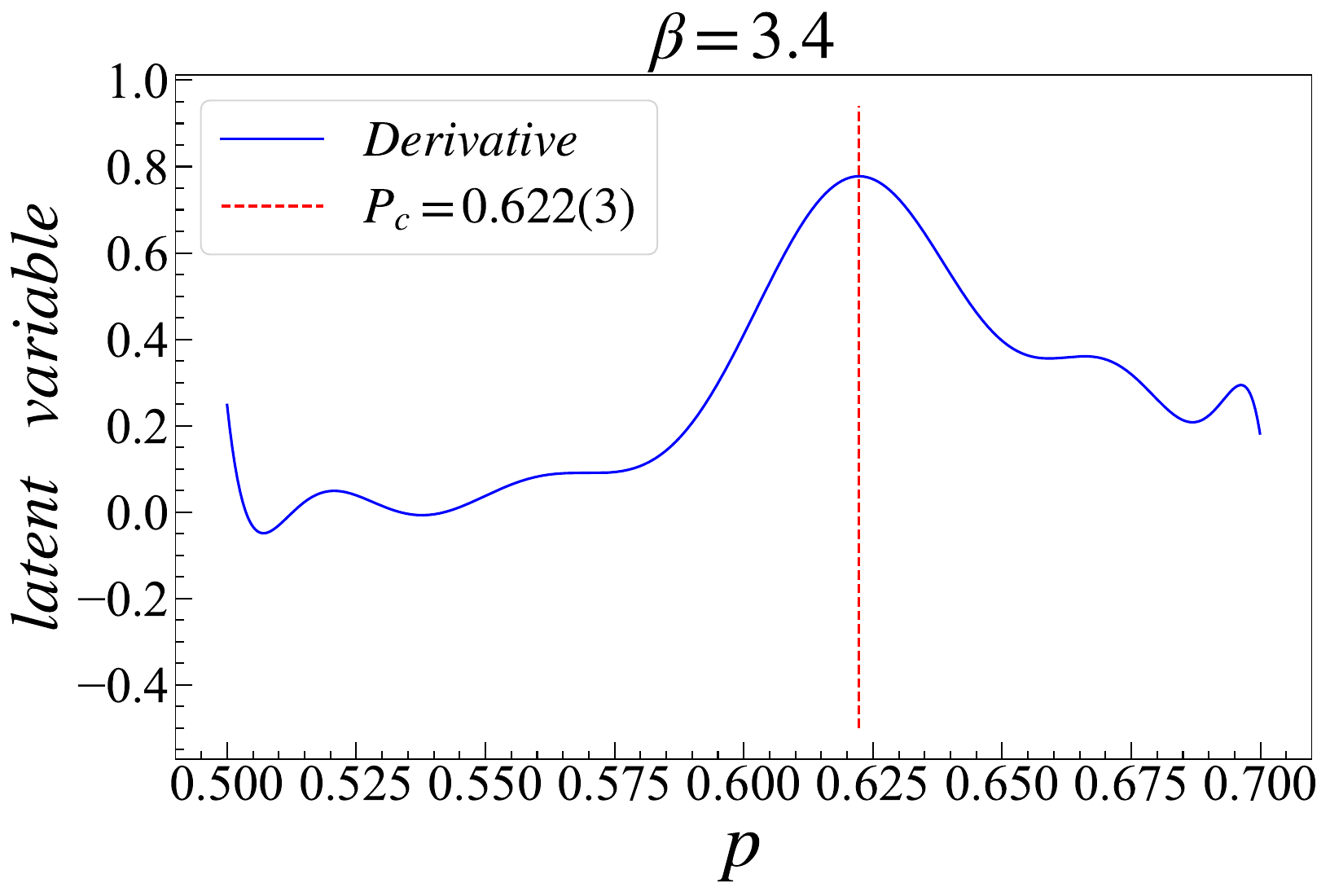} \\
    {\quad}{\quad}{\quad}(a) & $\qquad$ {\quad}{\quad}(b)
\end{tabular}
\caption
{At ${\beta}=3.4$, the one-dimensional encoding output of the hidden layer in SAEs, and the determination of the LDP critical point $P_c$. The system settings are consistent with those in Figure \ref{f_5}. Due to the non-uniqueness of the monotonicity of the fitted results of the one-dimensional encoding output of stacked autoencoders, with different values of ${\beta}$, the critical point can be determined based on the characteristics of extremum in its derivative curve. According to the global maximum value of extremum in Figure (b), the system's critical point is identified as $P_c=0.622(3)$.}
\label{f_6}
\end{figure*}

\begin{figure*}[t]
\begin{tabular}{cc}
    \includegraphics[width=0.49\textwidth]{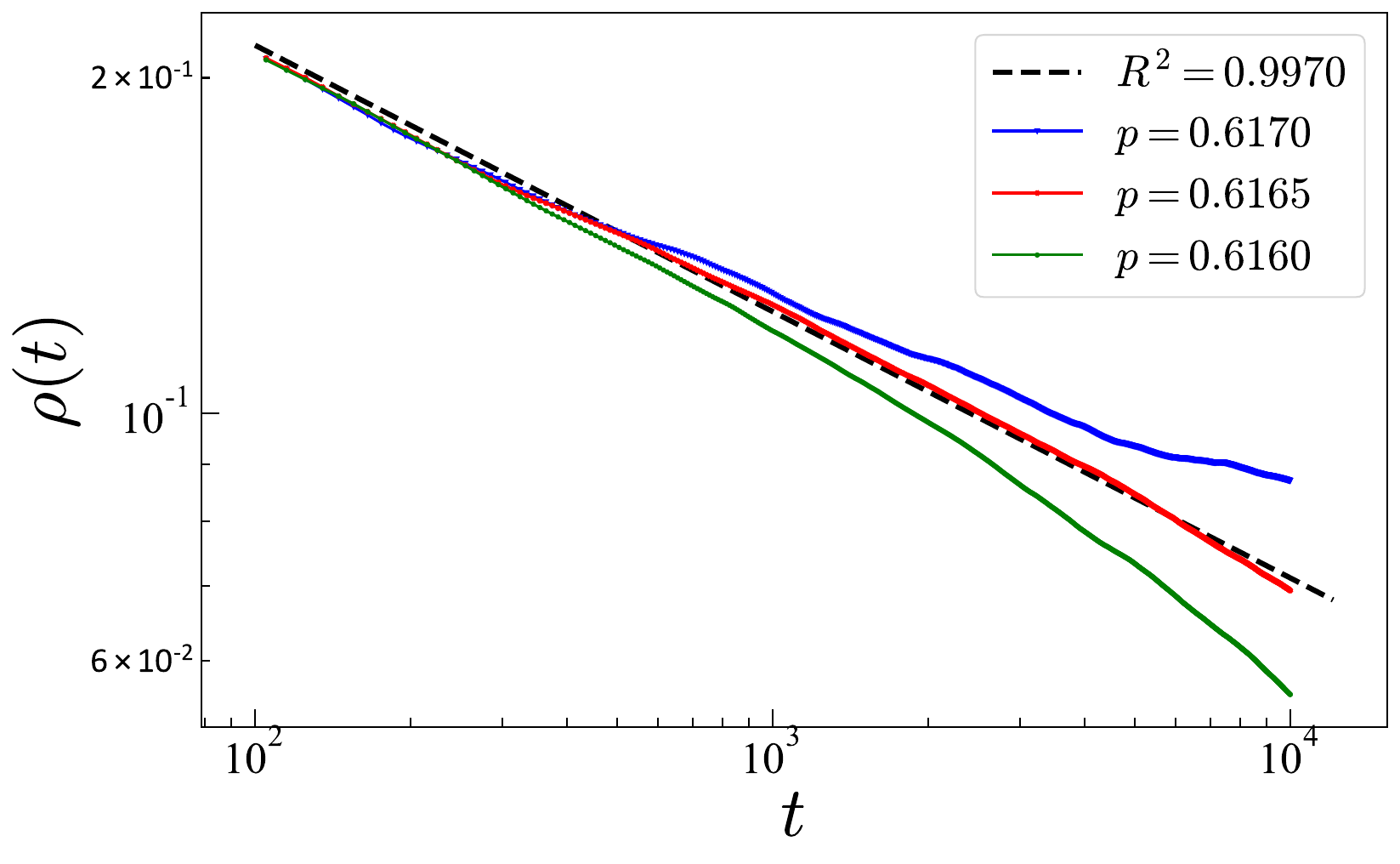} &
    $\qquad$\includegraphics[width=0.48\textwidth]{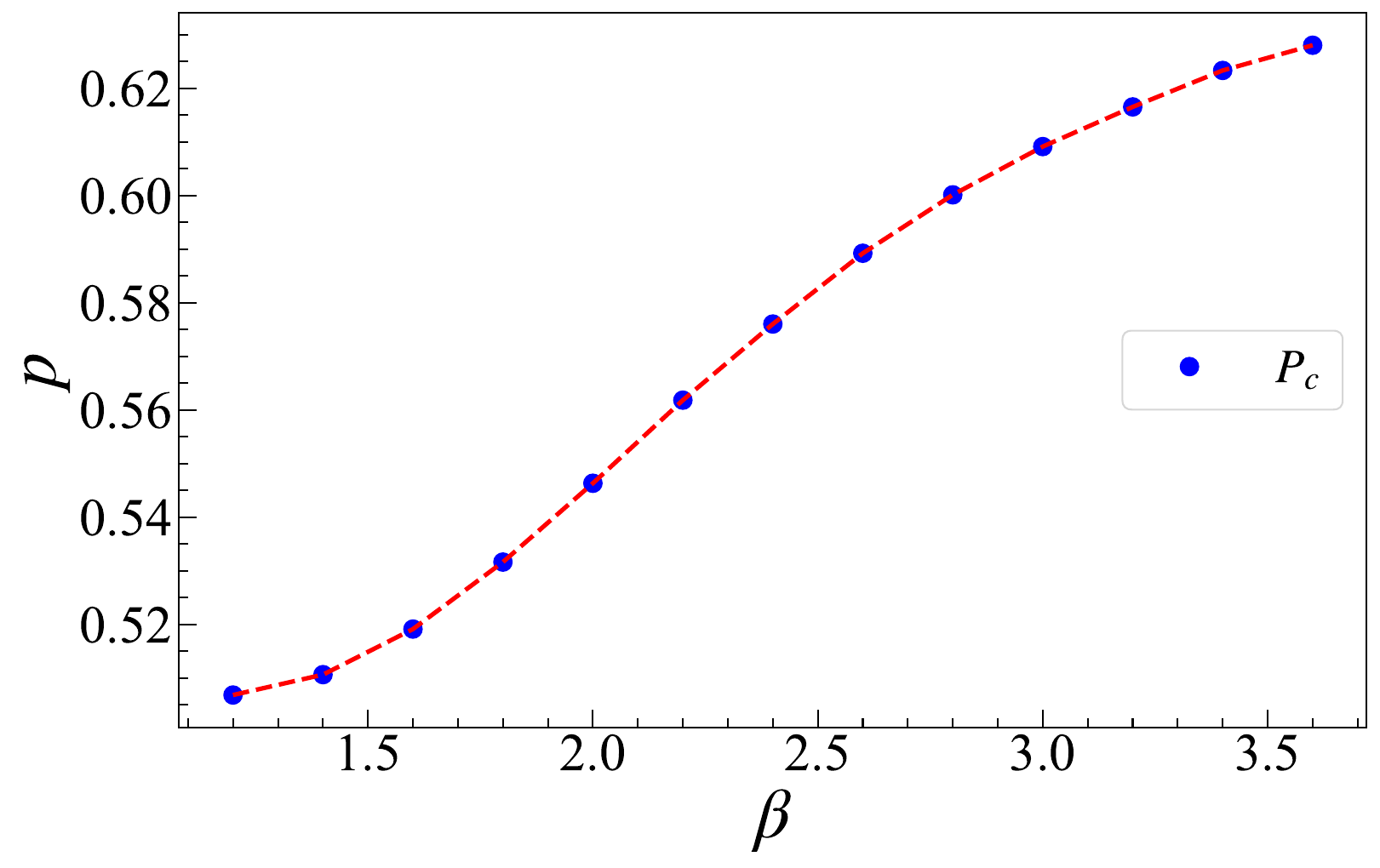} \\
    {\quad}(a) & $\qquad$ {\quad}{\quad}(b)
\end{tabular}
\caption
{(a) At ${\beta}=3.2$ and $p=0.6160,0.6165,0.6170$, the decay of the system's active particle density and the goodness of fit. The system size is $L=10000$, and the time step is set to $t=10000$. The black dashed curve represents the function curve fitted using a power-law, and the definition of goodness of fit is described in the text, with a best resulting $R^2=0.9970$. Therefore, the measurement error of this critical point from the true value is less than $0.0005$.(b) A set of critical points ${P_c}$ corresponding to different values of ${\beta}$. The variation of critical points corresponding to different values of ${\beta}$ (specific values are provided in the text). It can be inferred that the critical points undergo continuous changes under the control of the parameter ${\beta}$.}
\label{f_7}
\end{figure*}


We designed a SAE structure based on a fully connected neural network, and Figure \ref{f_3_1} illustrates the encoding and decoding processes of the autoencoder. We chose mean squared error (MSE) as the loss function to assess the data reconstruction capability of the autoencoder. During the encoding and decoding processes, we used multiple hidden layers to preserve the structural information of the initial data as much as possible and employed dynamic learning rates to optimize the model's parameter configuration. In the parameter updating process, we utilized the Adam optimizer to introduce momentum-corrected biases and introduced regularization to weaken training noise. Finally, we extracted the hidden variables encoded to the specified dimension.


Our basic workflow for using SAE generally includes the following steps. Firstly, we fix the value of the hyperparameter $\beta$. Cluster graphs with different transition probabilities, as shown in Figure \ref{f3}, are fed into the SAE, corresponding to the 'Input' layer in Figure \ref{f_3_1}. Similar to the left part of Figure \ref{f_3_1}, the SAE then encodes the original data through fully connected layers with gradually decreasing neurons and applies the ReLU activation function to provide non-linear mapping. After encoding, the original input data is dimensionally reduced, and the dimensionality is determined by specifying the number of neurons in the 'Encoder output' layer.

Decoding is the reverse process of encoding, aiming to reshape the original cluster graph using the low-dimensional results obtained through encoding. After multiple backpropagation and parameter updates, mean squared error is used as the loss function to evaluate the data reconstruction performance. Once SAE training is complete, we retain the encoding results with different numbers of neurons. The results of two neurons correspond to the positions of brown points in Figure \ref{f_3_1}, while the result of one neuron corresponds to the position of the red point in Figure \ref{f_3_1}. As shown in Figure \ref{f_4}, we use different colors to label the results of the two neurons. The result of one neuron is represented by a single color, such as the blue circle in Figure \ref{f_5}(a).

\section{Autoencoder and numerical results}
\subsection{Determination of critical points}
For absorbing phase transitions, the standard Monte Carlo method for identifying the critical point involves statistically analyzing the particle survival probability from a single particle source \cite{henkel2008non}. Due to limited prior knowledge, this typically requires starting from the boundaries and iteratively testing values between $0$ and $1$, gradually narrowing the possible range of the critical point using a bisection method. With a single particle source, the system size and evolution time steps often need to be significantly large, which not only increases computational demands but also places greater constraints on simulation details such as boundary conditions \cite{henkel2008non}. In contrast, unsupervised machine learning techniques, like stacked autoencoders, can effectively map physical quantities such as particle density during training while retaining information about the dynamic evolution of the system's spatial structure. Consequently, unsupervised learning based on stacked autoencoders has the potential to identify distribution differences across various states in smaller system sizes, allowing for more effective predictions of the critical point.

We initially employ an autoencoder to extract and analyze the two-dimensional features of cluster configurations from the LDP model. Considering training costs and precision, we select $41$ $p$ values at intervals of $0.005$ within the range of $[0.5,0.7]$. For each $p$, we repeatedly generate cluster configuration, resulting in a training set of $41{\times}500$ cluster configurations and a test set of $41{\times}50$ cluster configurations. The system size is\;$L=500$\;, the time step is set to \;$t=500$\;, and the value of the hyperparameter $\beta$ is chosen to be $3.2$ firstly. 
To maintain the stability of the stacked autoencoder's output, we established an adequate number of training epochs to ensure that accuracy and loss values stabilize.

To retain as much cluster diagram structural information as possible, we employ full-seed initial conditions. By setting the hidden layer to be two-dimensional, we extract the two-dimensional structural features of the cluster configuration, as illustrated in Figure \ref{f_4}. In the figure, $h1$ and $h2$ represent the coordinates of the two-dimensional feature points, and the color bar on the right indicates the corresponding colors for different branching probability values. The results indicate that near the critical point, the feature points are blurred and dispersed, even fractured, suggesting that the autoencoder has captured the particularities near the critical point of the system. We infer that the autoencoder can distinguish the power-law growth characteristics of particle density near the critical point and specific structural features of the cluster configuration from regions far from the critical point. Based on this idea, we set the hidden layer to be one-dimensional and identify the critical point through the features of the one-dimensional data.



We select ${\beta}=3.2,3.4$ and input the corresponding training and test sets into the autoencoder, obtaining the results of the autoencoder's extraction of one-dimensional features from the cluster diagrams, as shown in Figure \ref{f_5}(a) and Figure \ref{f_6}(a). Observing the features of the curves in Figure \ref{f_5}(a) and Figure \ref{f_6}(a), when $p\to P_c$, the curves appear to reach maximum curvature. Therefore, we perform a polynomial fitting of the curves and plot the relationship between the derivative of the fitted curve and $p$. The results are depicted in Figure \ref{f_5}(b) and Figure \ref{f_6}(b). We find that the global minimum or maximum in the extreme values in the figures corresponds to the transition probability, which can be used as our estimate for $P_c$. According to the results in Figure \ref{f_5}(b) and Figure \ref{f_6}(b), when ${\beta}=3.2$ and $3.4$, we estimate their corresponding critical points to be $P_c=0.616(5)$ and $0.622(3)$. 


To verify the reliability of this method, we statistically analyze the changes in particle density of the system when it is close to the transition probability. According to the characteristics of absorbing phase transitions, the decay form of particle density at the critical point follows 
\begin{equation}
 {\rho}(t){\sim}t^{-{\delta}}.
 \label{12}
\end{equation}




When the system is not at the critical point, it deviates from this characteristic. With a system size of $L=10000$ and a time step of $t=10000$, we repeated the measurement of particle density changes $100$ times and averaged the results. As illustrated in Figure \ref{f_7}(a), significant power-law deviations were observed at transition probabilities $p=0.6160$ and $0.6170$. To confirm the accuracy of these deviations, we conducted a power-law fitting and measured the goodness of fit.

\begin{table*}[!tbp]
	\centering
\resizebox{500pt}{8mm}{	
	\begin{tabular}{cccccccccccccc}
			\hline
			\;${\beta}$\; & 1.2 & 1.4 & 1.6 & 1.8 & 2.0 & 2.2 & 2.4 & 2.6 & 2.8 & 3.0 & 3.2 & 3.4 & 3.6\\
			\hline
			\;$P_c$\; & 0.506(8) & 0.510(6) & 0.519(1) & 0.531(6) & 0.547(0) & 0.561(8) & 0.576(0) & 0.589(2) & 0.600(0) & 0.609(1) & 0.616(5) & 0.622(3) & 0.628(0)\\
			\hline
		\end{tabular}
		}

\caption
{A set of critical points corresponding to $\beta$ was obtained through the analysis of the stacked autoencoder's one-dimensional encoding results.}
\label{table6_1}
\end{table*}

\begin{equation}
R^2=1-\frac{\sum\left(y_a-y_p)\right)^2}{\sum\left(y_a-y_m\right)^2},
\label{13}
\end{equation} 

where $y_a$ represents the statistically measured actual density, $y_p$ represents the predicted values corresponding to the fitted curve, and $y_m$ represents the mean value of the actual measurements.

When ${\beta}$ is $3.2$, the goodness of fit for the transition probabilities $p=0.6165, 0.6170, $ and $0.6155$ are $R^2=0.9970, 0.9886$, and $ 0.9819$, respectively. Given that the optimal goodness of fit value is $1$, we identify $p=0.6165$  as the critical point of the system. With a margin of error less than $0.0005$ based on the variance observed, we determine the critical point to be $P_c=0.616(5)$.



To investigate the relationship between the critical point and the hyperparameter ${\beta}$, we duly select a series ${\beta}$ to measure $P_c$. At ${\beta}=3.6,3.4,3.2,3.0,2.8,2.6,2.4,2.2$, we select 41 $p$ as transition probabilities at intervals of 0.005 within the range of $[0.5,0.7]$ to generate training and test sets. Similarly, at ${\beta}=2.0,1.8,1.6,1.4,1.2$, we choose 41 $p$ within the range of $[0.45,0.65]$. All the measurements of $P_c$ are summarized in Table \ref{table6_1}.

\begin{figure*}[t]
\begin{tabular}{cc}
    \includegraphics[width=0.45\textwidth]{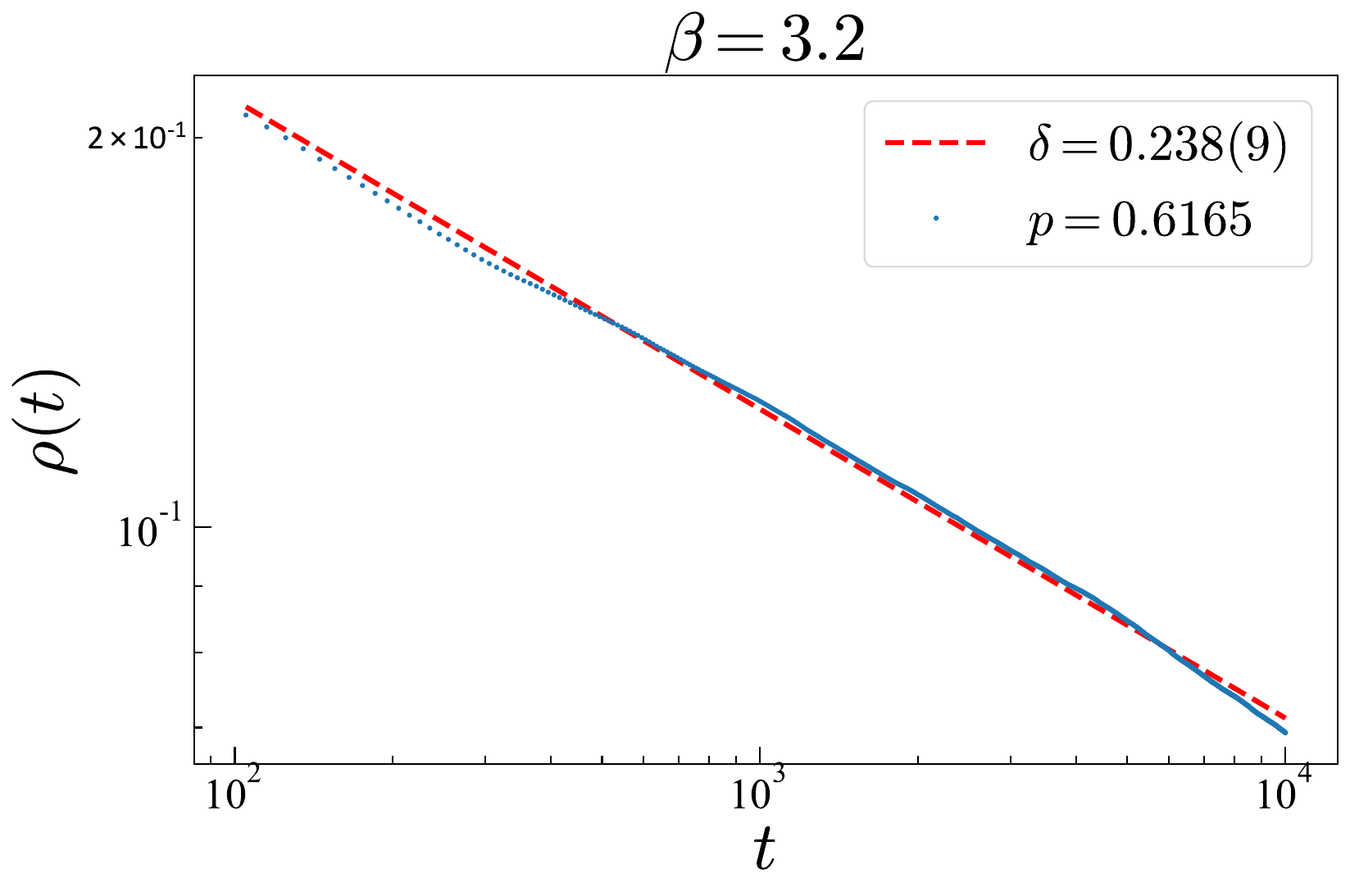} &
    $\qquad$\includegraphics[width=0.45\textwidth]{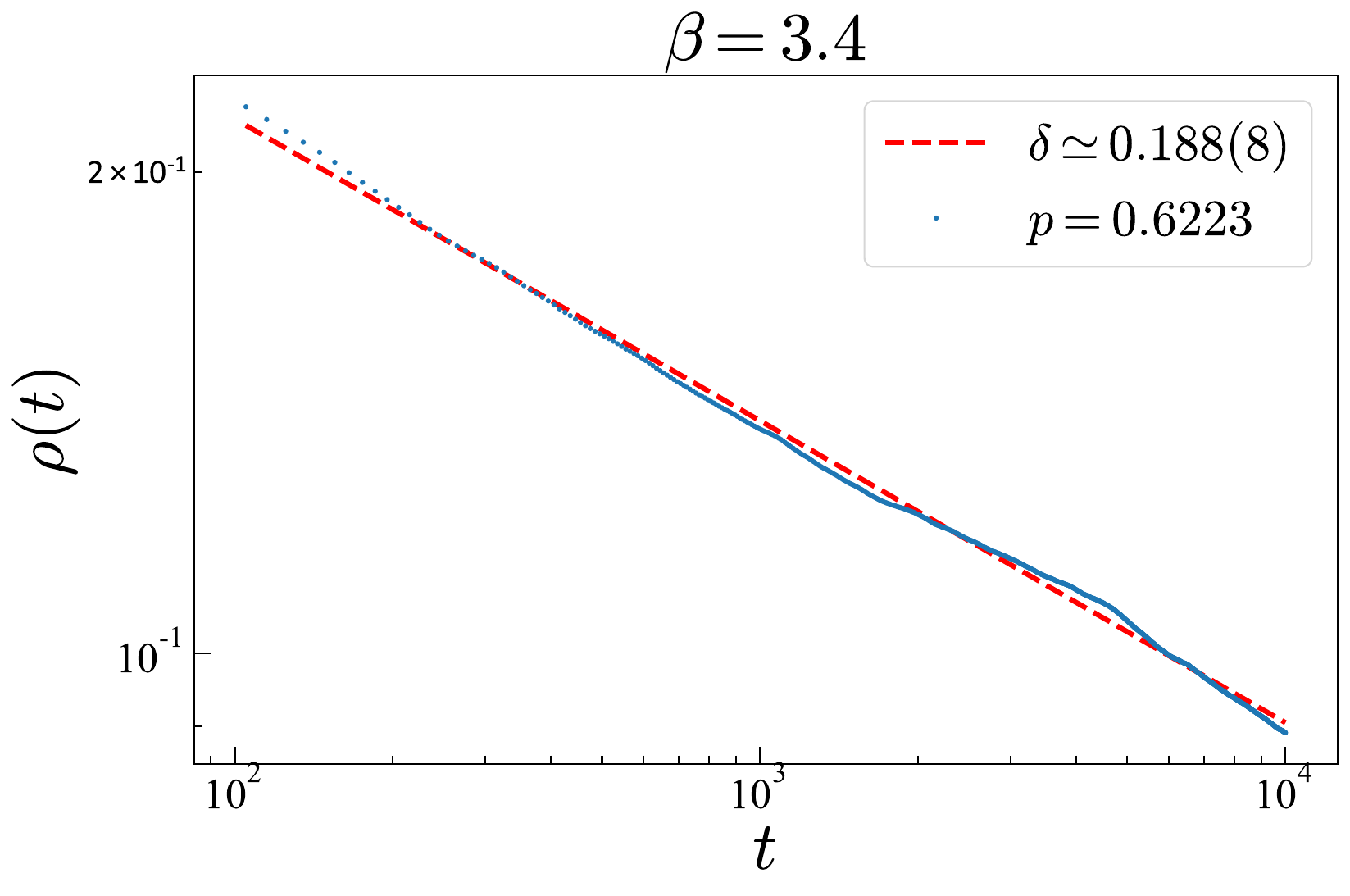} \\
    {\quad}(a) & $\qquad$ {\quad}{\quad}(b)
\end{tabular}
\caption
{The measurement of the critical exponent ${\delta}$ is conducted with the system settings identical to those in Figure \ref{f_7}. (a) At ${\beta}=3.2$, we present the decay of active particle density at the critical point $P_c=0.616(5)$ in a double-logarithmic coordinate system. According to Equation \ref{12}, the critical exponent is estimated as ${\delta}{\simeq}0.238(9)$. (b) At ${\beta}=3.4$, the value of the critical exponent ${\delta}$ at the critical point $P_c=0.622(3)$ is determined to be 0.188(8).}
\label{f_8}
\end{figure*}

\begin{table*}[!tbp]
	\centering
\resizebox{500pt}{8mm}{	
	\begin{tabular}{cccccccccccccc}
			\hline
			\;${\beta}$\; & 1.2 & 1.4 & 1.6 & 1.8 & 2.0 & 2.2 & 2.4 & 2.6 & 2.8 & 3.0 & 3.2 & 3.4 & 3.6\\
			\hline
			\;${\delta}$\; & 0.633(1) & 0.602(9) & 0.542(7) & 0.517(1) & 0.483(9) & 0.411(3) & 0.342(1) & 0.272(5) & 0.282(5) & 0.253(7) & 0.238(9)4 & 0.188(8) & 0.173(8)\\
			\hline
		\end{tabular}
		}

\caption
{A set of critical points corresponding to $\beta$ and the critical exponent $\delta$ obtained through statistical measurements of the system's active particle density.}
\label{table6}
\end{table*}

\begin{figure*}[t]

\begin{tabular}{cc}
    \includegraphics[width=0.45\textwidth]{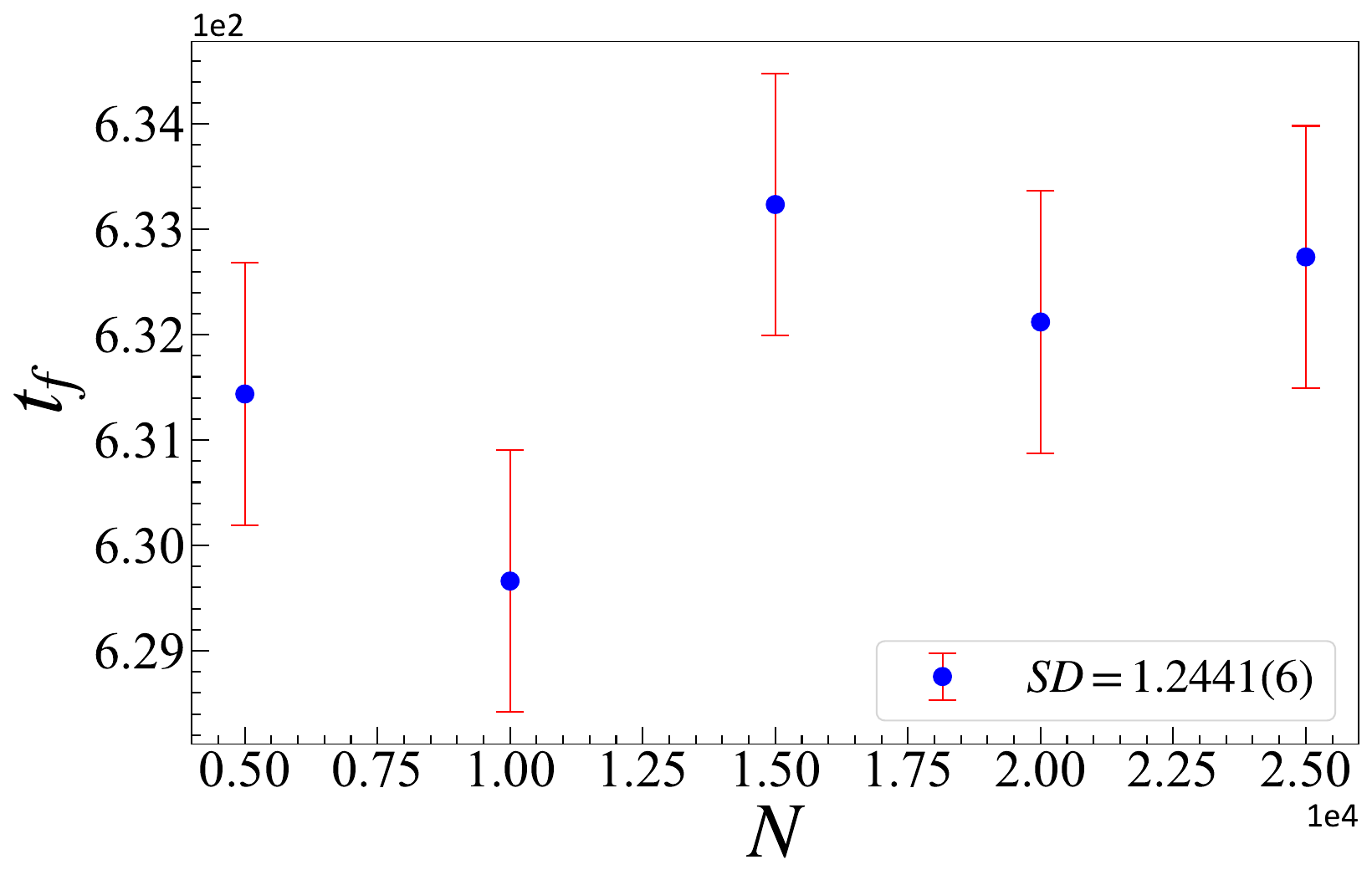} &
    $\qquad$\includegraphics[width=0.45\textwidth]{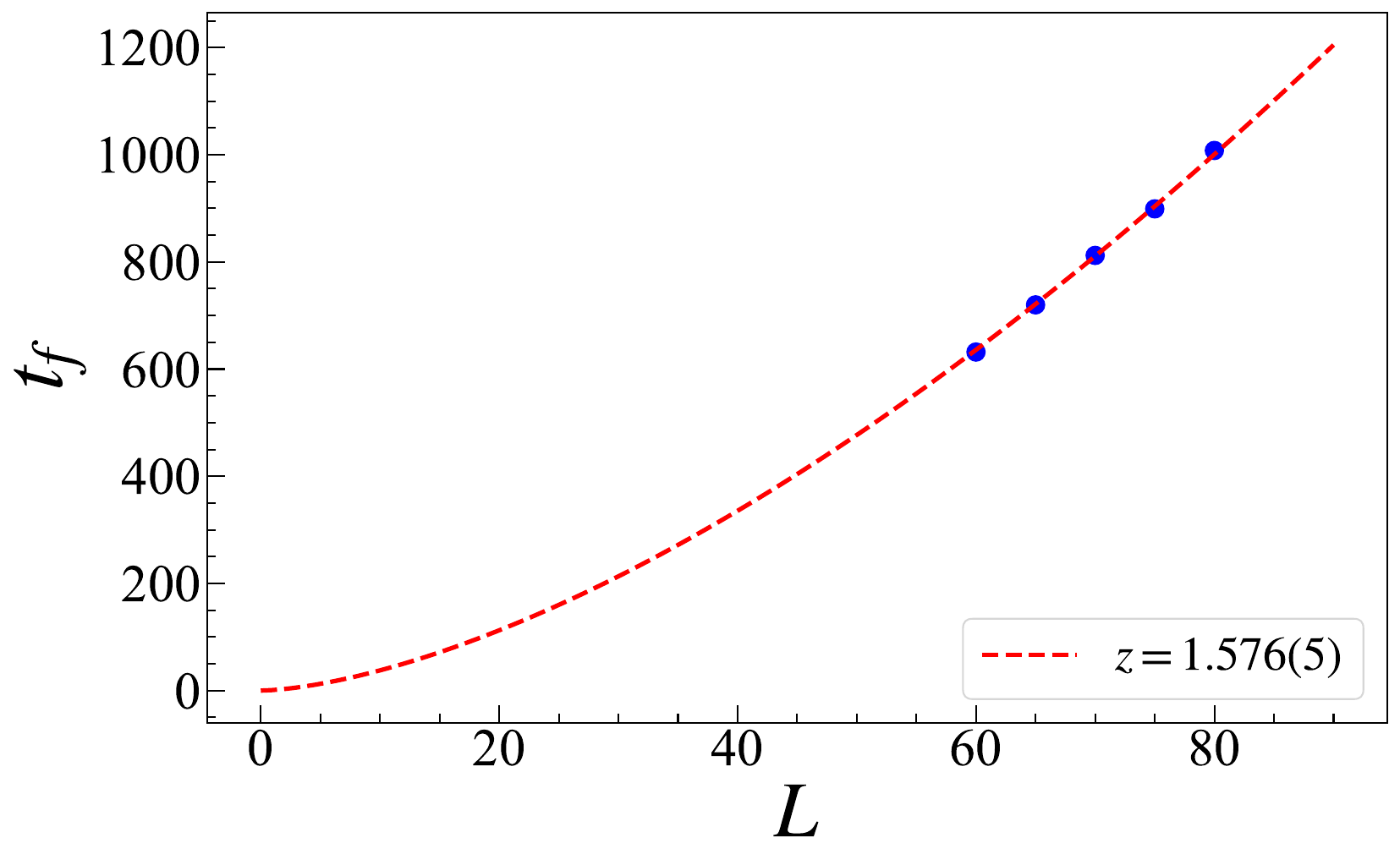} \\
    {\quad}{\quad}(a) & $\qquad$ {\quad}{\quad}(b)
\end{tabular}

\caption
{
At ${\beta}=3.6$, the standard deviation of characteristic times for systems with the same size and the measurement of the critical exponent $z$ for systems with different sizes. (a) For $L=60$ and $t=1200$, the average values of $t_f$ taken every $5000$ steps are computed, and the standard deviation of these five measurements is calculated as $SD=1.2441$. (b) System sizes $L=60,65,70,75,80$ with corresponding time step settings of $1200,1400,1600,1800,2000$, respectively, are used to calculate the statistical values of characteristic times, with each $t_f$ being the result of averaging over $25000$ statistical measurements. Utilizing the relationship (\ref{16}) between characteristic times and dynamic exponent $z$, the critical exponent is determined as $z=1.576(5)$.
}
\label{f_11}
\end{figure*}

\subsection{Measurement of critical exponent \texorpdfstring{$\delta$}{}}


In the above work, we are also able to estimate the density decay exponent $\delta$ of the system particles. To reduce statistical errors, we measured the particle density decay of larger-sized systems based on the critical points determined by the autoencoder. Specifically, we initially observed the variation in particle density for system sizes of $L=100000$ and time steps of $t=10000$ at ${\beta}=3.2,3.4$ with $P_c=0.6165,0.6223$. To mitigate the impact of random errors, we averaged the results of the evolution with $100$ different initial conditions. As an example, the results of $\delta$ obtained through linear fitting in double-logarithmic coordinates are depicted in Figure \ref{f_8}. The fitting result for Figure \ref{f_8}(a) is $0.64311451t^{-0.0.23899697}$, and for Figure \ref{f_8}(b) it is $0.51525595t^{-0.18883931}$. Subsequently, we measured the values of the density decay exponent $\delta$ at  ${\beta}=3.6,3.0,2.8,2.6,2.4,2.2,2.0,1.8,1.6,1.4,1.2$ in Table \ref{table6}.

Regarding the measurement of critical points in Table \ref{table6_1}, we obtained different values of $P_c$ for various ${\beta}$, suggesting that $P_c$ in the LDP model continuously varies based on the hyperparameter $\beta$. The variation of the critical exponent $\delta$ indicates that, under the control of $\beta$,  the system deviates from the universality class of ordinary DP. This demonstrates that the introduction of long-range interactions alters the symmetry of the ordinary DP system.

\begin{figure*}[t]

\begin{tabular}{cc}
    \includegraphics[width=0.45\textwidth]{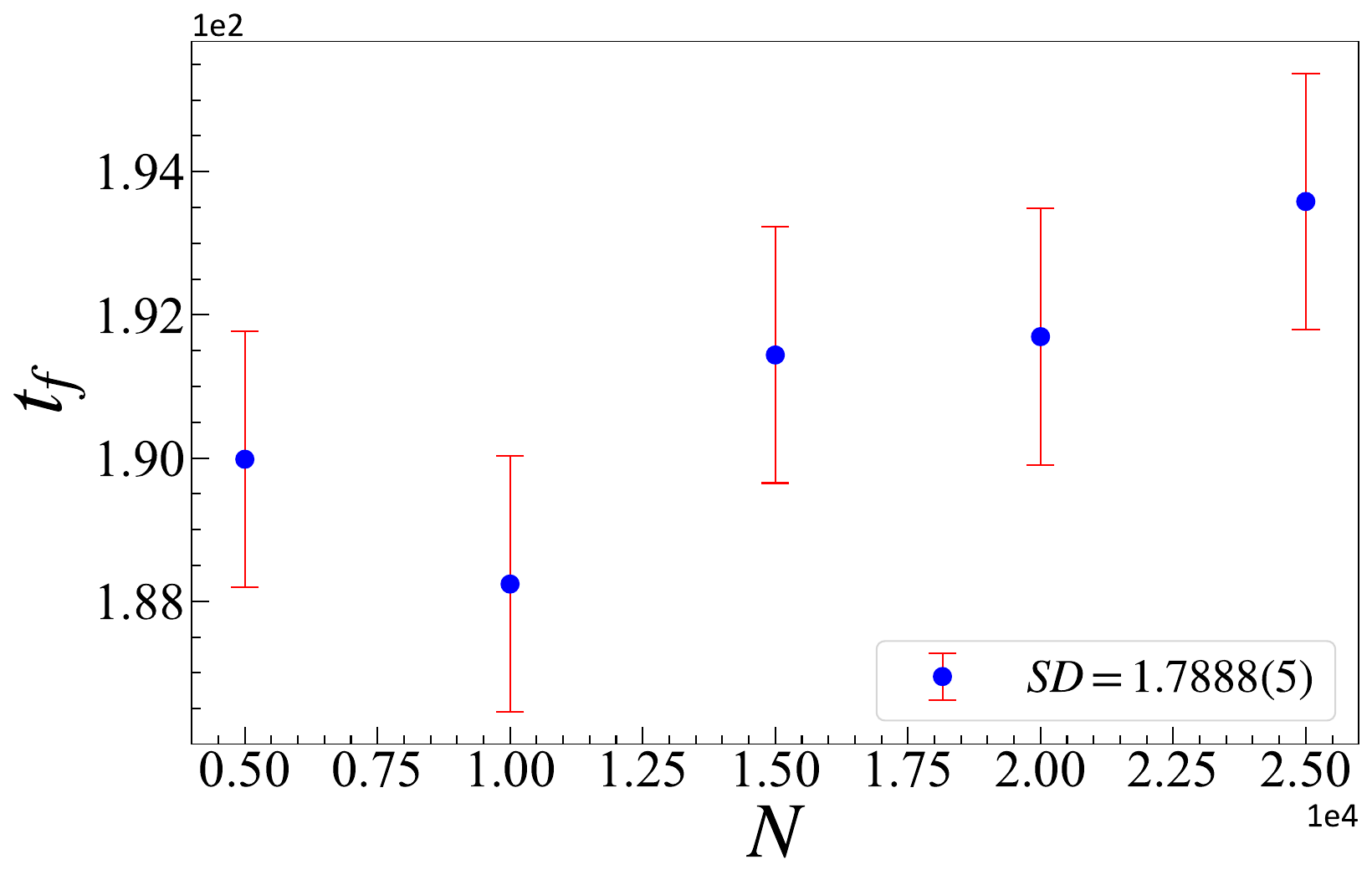} &
    $\qquad$\includegraphics[width=0.45\textwidth]{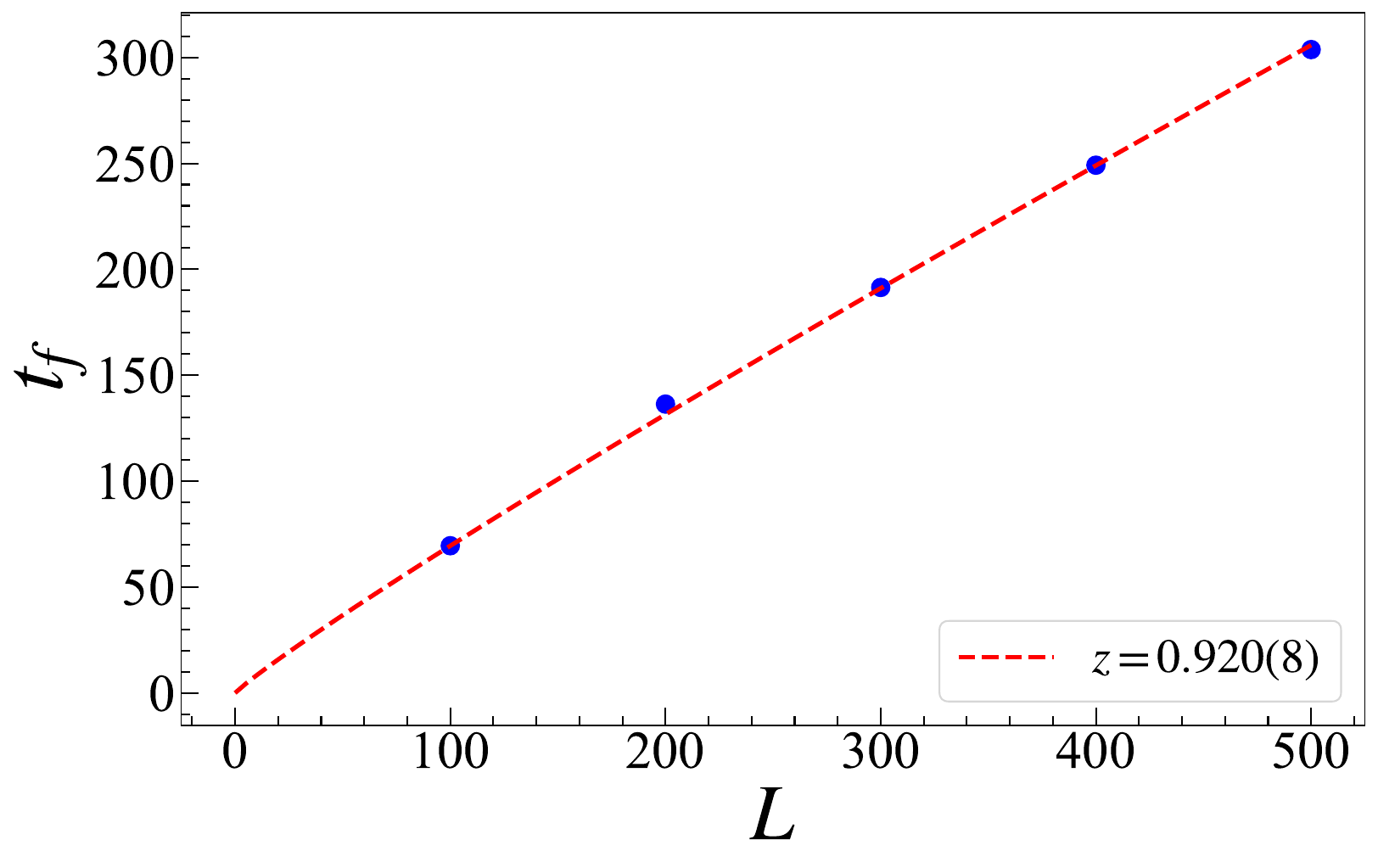} \\
    {\quad}{\quad}(a) & $\qquad$ {\quad}{\quad}(b)
\end{tabular}
\caption
{
At ${\beta}=1.8$, the standard deviation of characteristic times for systems with the same size and the measurement of the critical exponent $z$ for systems with different sizes. (a) For $L=300$ and $t=600$, the average values of $t_f$ taken every $5000$ steps are calculated, and the standard deviation of these five measurements is computed as $SD=1.7888$. (b) The statistical values of characteristic times for system sizes $L=100,200,300,400,500$ with corresponding time step settings of $200,400,600,800,1000$, respectively, are shown. Each $t_f$ represents the result of averaging over $25000$ statistical measurements. The fitted result for the critical exponent $z$ is $z=0.920(8)$.
}
\label{f_12}
\end{figure*}

\begin{figure*}[t]
\begin{tabular}{cc}
    \includegraphics[width=0.43\textwidth]{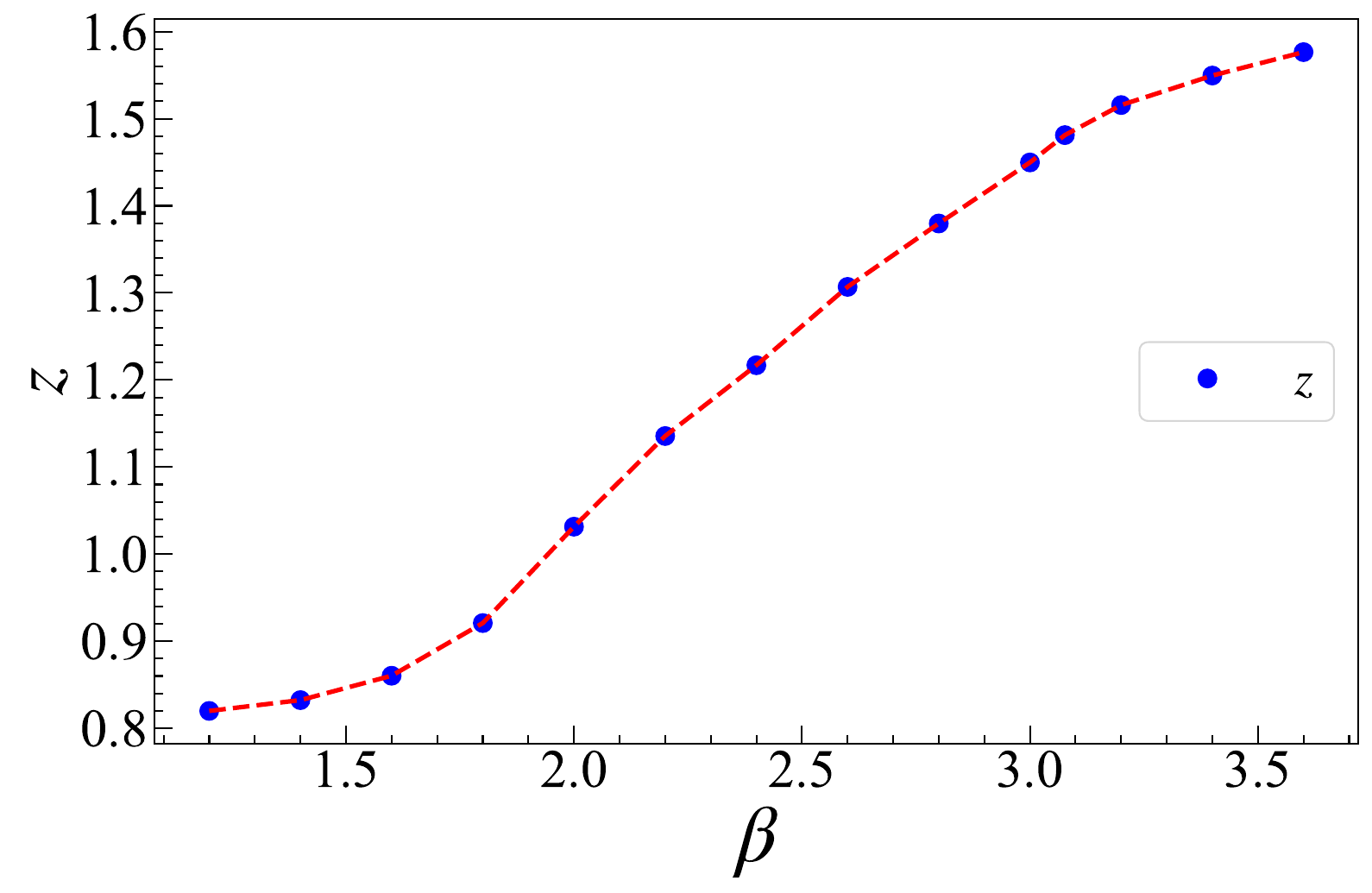} &
    $\qquad$\includegraphics[width=0.46\textwidth]{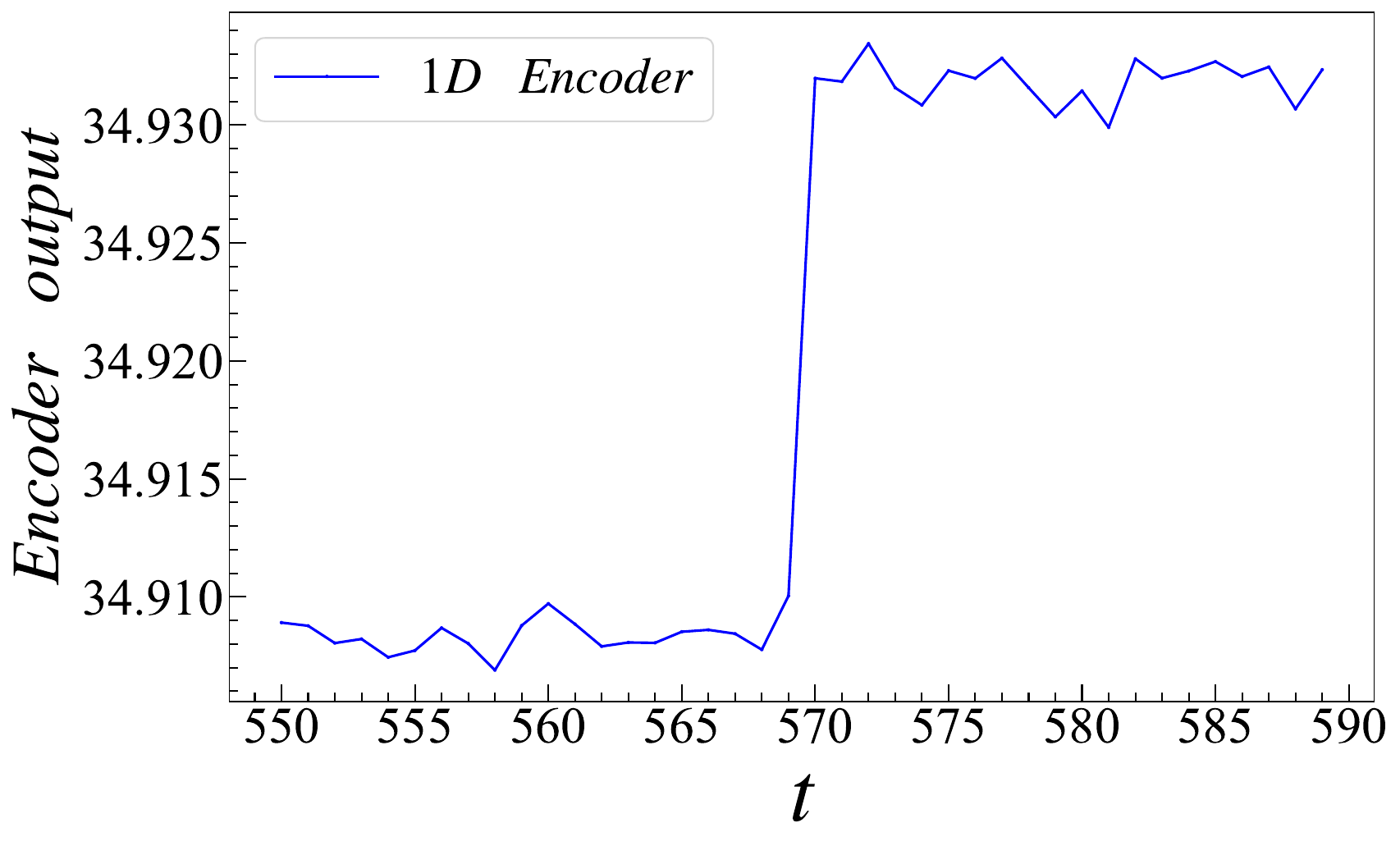} \\
    {\quad}{\quad}(a) & $\qquad$ {\quad}{\quad}(b)
\end{tabular}
\caption
{
(a) Measurements of the dynamic exponent $z$ corresponding to a set of ${\beta}$ values (specific values are provided in the text) suggest that the dynamic exponent continuously changes with the parameter ${\beta}$. Combining the measurements of the critical exponents ${\delta}$ and ${\Theta}$ mentioned above, indicates that the universality class to which the DP system belongs under spatial L{\'{e}}vy-like long-range interactions dynamically changes with the parameter ${\beta}$.
(b) At ${\beta}=3.4$ and $P_c=0.622(3)$, for a system of size $L=60$, the one-dimensional encoding output after training with stacked autoencoders for cluster plots at different time steps is presented. The presence of larger gaps in the plot, particularly those close to the predicted characteristic time $t_f=569.40$, indicates that stacked autoencoders can effectively recognize this evolutionary feature of the system.
}
\label{f_13}

\end{figure*}

\begin{figure}
\centering
\includegraphics[width=0.49\textwidth]{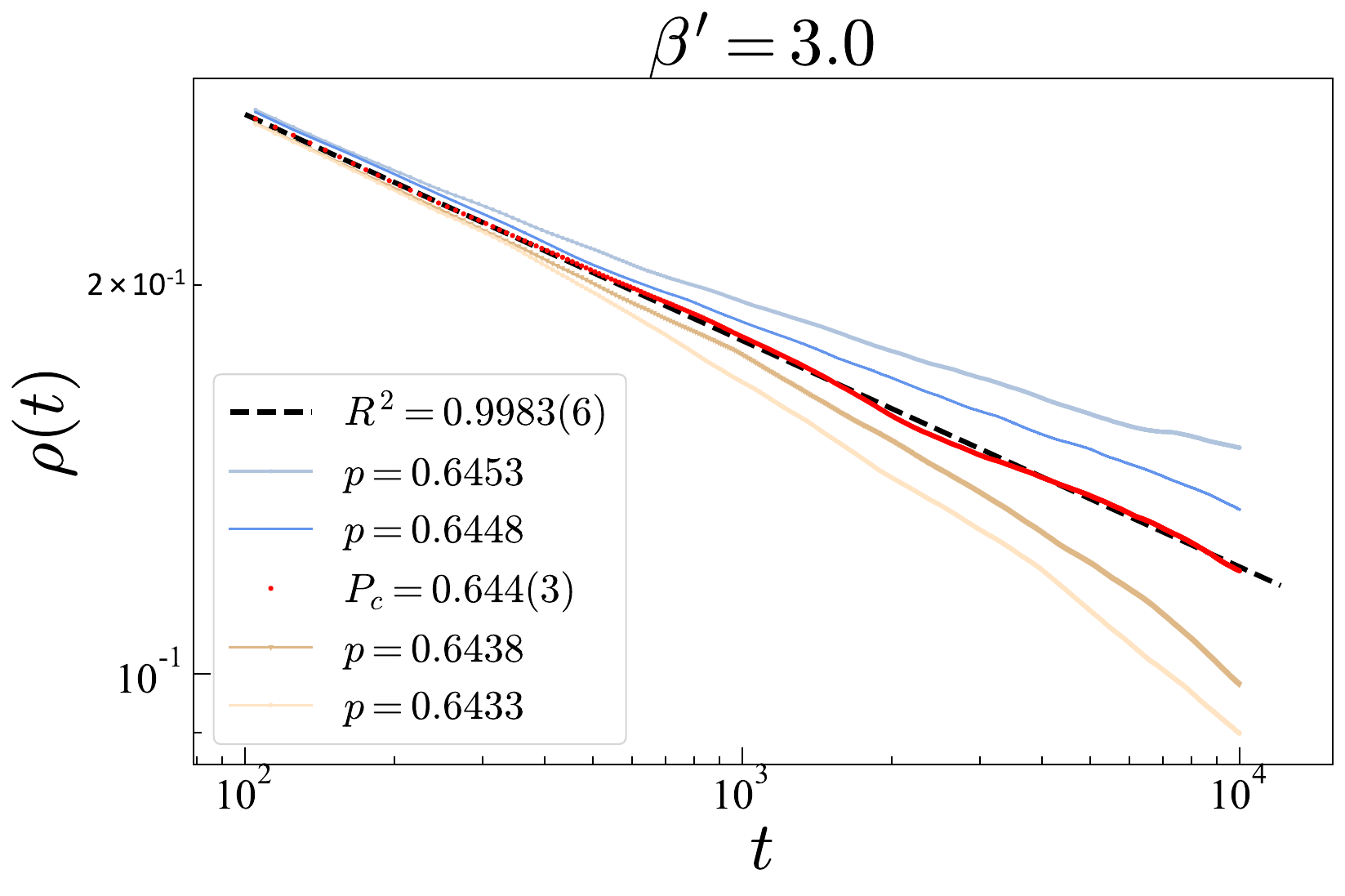}
\caption
{At ${\beta}^{\prime}=3.0$, the measurement of the critical point is conducted by introducing a global expansion mechanism using the L{\'{e}}vy distribution to generate random walk step lengths. 
Among the probabilities $p=0.6443,0.6443\pm0.0005$, and $0.6443\pm0.0010$, the system exhibits the best power-law fitting when $P_c=0.6443$, with maximum goodness of fit of $R^2=0.9983$. For probabilities far from this transition point, the particle density shows increasing deviations of the power-law, indicating a significant departure from the critical behavior.
}
\label{f_16}
\end{figure}

\subsection{Measurement of the dynamic exponent \texorpdfstring{$z$}{}}

To further validate the properties associated with the DP universality class in the context of long-range interactions, we aim to determine the system's dynamic exponent $z$. The dynamic exponent $z$ follows the scaling relation at the critical point. 
\begin{equation}
 {\xi}_{\perp}{\sim}t^{\frac{1}{z}}.
 \label{14}
\end{equation} 

And ${\xi}_{\perp}$ refers to spatial correlation length. Considering the relation between dynamical exponent $z$ and mean square spreading exponent $\widetilde{z}$, where $\widetilde{z} = 2/z$, and $\widetilde{z}$ satisfies the scaling relation
\begin{equation}
 r^2(t){\sim} t^{\widetilde{z}}.
\label{15} 
\end{equation} 




It is common to measure mean square spreading $r^2$ of surviving clusters from origin in place of the dynamical exponent $z$. However, simulations have shown that $r^2$ does not exhibit a power-law but diverges at the critical point in the presence of long-range interaction\cite{1998A}. In a finite lattice system, there is a non-vanishing probability of reaching the absorbing configuration. In the critical region, when the spatial size of the system is $L$, the system reaches an absorbing state after a characteristic time $t_f$, which satisfies the relation: 
\begin{equation}
 t_f{\sim}L^z.
\label{16} 
\end{equation}



We utilize finite-size effects to determine the dynamic exponent $z$. In brief, we record the time steps required for systems of multiple sizes to reach an absorbing configuration at critical probability. We then perform a fitting on those time steps to determine $z$.




Due to computational limitations, the upper time step limit set in our simulations is not smaller than the characteristic time of ordinary DP universality classes. The introduction of long-range interactions leads to the clustering structure becoming more discrete, and the absorbing state appearing more quickly. Compared to ordinary DP, such systems have a smaller dynamic exponent when $\beta$ is small. We selected five system sizes ($L=60,65,70,75,80$) for each value of ${\beta}=2.2,2.4,2.6,2.8,3.0,3.2,3.4,3.6$ and performed finite-size scaling analysis with a temporal scale limit of $1200,1400,1600,1800,2000$ for each system size. To reduce random errors, we performed five times ensemble averages for a total of $25000$ systems for each system size. Figure \ref{f_11}(a) displays the characteristic time $t_f$ and its standard deviation obtained from ensemble averages of systems with $\beta = 3.6$ and $L=60$. The small standard deviation can serve as a reference for measuring accuracy. Based on the power-law relationship between the characteristic time and finite size, we fitted the $t_f$ for different system sizes and calculated the dynamical exponent $z=1.5765(4)$, as shown in Figure\ref{f_11}(b). 



We find that both $z$ and $t_f$ decrease as $\beta$ decreases. To distinguish the measured values of $t_f $ for different finite sizes and reduce the system error, we increased the system's spatial scale($L=100,200,300,400,500$) for ${\beta}=1.2,1.4,1.6,1.8,2.0$. At the same time, to reduce simulation costs, we decreased the temporal scale limit of the system($t=200,400,600,800,1000$). In Figure \ref{f_12}, we show the measured value of $t_f$ and the fitted value of $z$ for ${\beta}=1.8,L=300,t=600$. Figure \ref{f_13}(a) displays the curve of $z$ for $\beta$. The dynamical exponents for other values of $\beta$ are listed in Table\ref{table7}.

\begin{table*}[!tbp]
	\centering
\resizebox{500pt}{8mm}{	
	\begin{tabular}{cccccccccccccc}
			\hline
			\;${\beta}$\; & 1.2 & 1.4 & 1.6 & 1.8 & 2.0 & 2.2 & 2.4 & 2.6 & 2.8 & 3.0 & 3.2 & 3.4 & 3.6\\
			\hline
			\;$z$\; & 0.819(9) & 0.832(3) & 0.860(3) & 0.920(8) & 1.031(5) & 1.13(5) & 1.216(9) & 1.306(9) & 1.379(7) & 1.449(8) & 1.515(7) & 1.549(6) & 1.576(5)\\
			\hline
		\end{tabular}
		}

\caption
{By statistically analyzing characteristic times for systems of different sizes but the same $\beta$, and fitting to obtain the values of the dynamic exponent $z$.}
\label{table7}
\end{table*}





After obtaining the above results, we used the cluster diagrams generated at different time scales as both training and testing inputs for an autoencoder. Our aim was to observe the one-dimensional hidden layer output in order to capture the critical evolution characteristics of the system. For example, under the conditions of ${\beta}=3.4$ and $P_c=0.6223$, we generated cluster configurations at different temporal scales for system size $L=60$, with the temporal scale ranging from $t{\in}[550,590]$. We generate a total of $41{\times}500$ cluster configurations as the training set and select $1/10$ of them as the testing set input into the autoencoder. Figure \ref{f_13}(b) demonstrates the results of the single latent layer variables. It is evident that the latent variable increases suddenly near the characteristic time, indicating that the autoencoder is also capable of recognizing critical features of the system.

\subsection{A new expansion mechanism}

To evaluate the universality of our approach, we measured the critical point for a distinct type of spatial long-range interaction within the DP system. Considering the specific math form of the Lévy distribution, we tried a method of generating a random walk step length that conforms to the Lévy distribution \cite{2010Nature}. The step length $s$ is generated by the following equation
\begin{equation}
 s=\frac{u}{|v|^{1/({\beta^{\prime}-1})}},
 \label{18}
\end{equation}
where $u,v$ follow normal distribution
\begin{equation}
 u{\sim}N(0,{\sigma}_u^2),{\quad}{\quad}v{\sim}N(0,{\sigma}_v^2).
 \label{19}
\end{equation}
Besides,
\begin{equation}
\sigma_u=\left\{\frac{\Gamma(\beta^{\prime}) \sin (\pi (\beta^{\prime} -1) / 2)}{\Gamma(\beta^{\prime} / 2) (\beta^{\prime}-1) 2^{(\beta^{\prime}-2) / 2}}\right\}^{1 / (\beta^{\prime} -1)}, \quad \sigma_v=1.
\label{20}
\end{equation}



Following the outlined rules, we generated $2000$ random step lengths for ${\beta}^{\prime}=2.0$ and ${\beta}^{\prime}=3.0$
The generated step sizes obtained by the above method follow a symmetrical Levy stable process with distribution\cite{mantegna1994fast}
\begin{equation}
L_{\alpha, \gamma}(z)=\frac{1}{\pi} \int_0^{\infty} \exp \left(-\gamma q^\alpha\right) \cos (q z) d q.
\end{equation}
Here, parameters $\alpha$ and $\gamma$ are used to characterize the features of the distribution function. Specifically, parameter $\alpha ={\beta}^{\prime}-1$ affects the properties of the distribution and the scaling characteristics of the random process, while parameter $\gamma$ is used to define the scale unit of the distribution function.

We replaced the step sizes $L$ and $R$ with $s$ in Sec.\uppercase\expandafter{\romannumeral3}. A, and measured the critical point of the system, at ${\beta}^{\prime}=3.0$, to be $P_c=0.644(3)$. In Figure \ref{f_16}, it can be observed that when the transition probability deviates from $P_c=0.644(3)$, the changes in particle density exhibit different degrees of deviation in the power-law behavior. This phenomenon indicates that our method is well-suited for DP systems with different spatial long-range interactions. We further speculate that the critical point determination method, developed using a stacked autoencoder, shows potential for broader application in other absorbing phase transition systems involving long-range interactions.




\section{Conclusion}


We studied the (1+1)-dimensional DP model with power-law distributed spatial long-range interactions using stacked autoencoders and MC methods. By varying the hyperparameter $\beta$, we determined the system's critical points and measured several critical exponents.





First, we designed a spatial long-range interaction DP evolution program based on the power-law random walk steps. The results of the cluster diagram indicate that spatial long-range interactions alter the ordered structure of the system, enhancing the influence of fluctuations and thereby modifying the upper critical dimension of the model. Using a SAE, we identified the critical points of the system controlled by the hyperparameter $\beta$. Utilizing the one-dimensional encoding results from the stacked autoencoder, our analysis revealed that the extrema, both maxima and minima, of the derivative of the curve serve as effective indicators for identifying critical points. Furthermore, we performed a statistical examination of the decay behavior of the particle density in proximity to the critical point. Through a comparative assessment of the variations in particle density across different transition probabilities, alongside an evaluation of their goodness of fit, we substantiated the robustness of this methodology in accurately pinpointing critical points. Subsequently, we record the variations in critical points of the system corresponding to different $\beta$ values, as shown in Figure \ref{f_7}(a) and Table \ref{table6_1}. We infer that in the DP system with such spatial long-range interactions, the critical point $P_c$ undergoes continuous changes with the parameter $\beta$.

To explore the universality class to which the DP system with such L\'evy-like spatial long-range interactions belongs, we measured the critical exponents ${\delta}$ and $z$ of the system. 
Our findings suggest that the universality class of the DP system varies with the hyperparameter $\beta$ under L\'evy-like flight interactions. Notably, by using clusters at different evolutionary time steps as training data, we observed that the one-dimensional encoding results from the stacked autoencoder effectively identify the characteristic time. 
This implies that stacked autoencoders may have more applications in measuring critical exponents. 

Finally, we tried a new step length generation algorithm, where step lengths obey the L\'evy distribution. 
The results demonstrate that our method still accurately determines the critical point, highlighting the versatility of our approach.
Additionally, introducing different forms of long-range interactions in space and time may broaden the practical application range of the DP evolution mechanism. 
We believe that unsupervised learning algorithms, particularly those based on autoencoder structures, hold promising potential for applications in such reaction-diffusion processes.


\section{Acknowledgements}
This work was supported in part by the Key Laboratory of Quark and Lepton Physics (MOE), Central China Normal University(Grant No.QLPL2022P01), the National Natural Science Foundation of China (Grant No. 11505071, 61702207 and 61873104), and the 111 Project, with Grant No. BP0820038.

\nocite{*}

\bibliography{apssamp}

\end{document}